\documentclass[12pt]{article}
\usepackage{a4wide}
\usepackage{graphicx}
\usepackage{amsfonts}
\usepackage{amstext}
\usepackage{amsmath}
\usepackage{amssymb}
\usepackage{amsthm}
\usepackage[mathscr]{eucal}
\makeatletter
 
 \@addtoreset{equation}{section}
\newtheorem{theorem}{Theorem}
\newtheorem{proposition}[theorem]{Proposition}
\newtheorem{lemma}[theorem]{Lemma}
\newtheorem{definition}[theorem]{Definition}
\newtheorem{corollary}[theorem]{Corollary}
\@addtoreset{theorem}{section}

\makeatother
\def\a{\alpha}
\def\A{\mathcal{A}}
\def\Ai{{\rm Ai}}
\def\b{\beta}

\def\B{\mathcal{B}}
\def\C{\mathbb{C}}
\def\d{\delta}

\def\e{\epsilon}
\def\E{\mathbb{E}}
\def\g{\gamma}

\def\GG{\mathbb{G}}
\def\i{\infty}

\def\l{\lambda}

\def\o{\omega}
\def\O{\Omega}

\def\vp{\varphi}
\def\P{\mathbb{P}}
\def\1{\bf{1}}
\def\R{\mathbb{R}}

\def\S{\mathcal{S}}

\def\t{\tau}
\def\th{\theta}
\def\T{\mathbb{T}}
\def\th{\theta}

\def\Z{\mathbb{Z}}
\def\Zm{\mathcal{Z}}
\def\z{\zeta}

\def\k{\kappa}

\begin{document}
\title{Free energy distribution of the stationary O\rq{}Connell-Yor directed random polymer model}
\author{Takashi Imamura
\footnote { Department of Mathematics and Informatics, 
Chiba University,~E-mail:imamura@math.chiba-u.ac.jp}
, Tomohiro Sasamoto
\footnote { Department of Physics, 
Tokyo Institute of Technology,~E-mail: sasamoto@phys.titech.ac.jp}}

\maketitle

\begin{abstract}
We study the semi-discrete directed polymer model introduced by O\rq{}Connell-Yor 
in its stationary regime, based on our previous work on the stationary $q$-totally 
asymmetric simple exclusion process ($q$-TASEP) using a two-sided $q$-Whittaker 
process. We give a formula for the free energy 
distribution of the polymer model in terms of Fredholm determinant and show that 
the universal KPZ stationary distribution appears in the long time limit. We also 
consider the limit to the stationary KPZ equation and discuss the connections with 
previously found formulas. 
\end{abstract}

\section{Introduction}
The O\rq{}Connell-Yor (OY) polymer model introduced in \cite{OY2001} is 
a finite temperature directed polymer model in a Brownian motion environment. 
At zero temperature, this is related to the GUE random matrix 
\cite{Baryshnikov2001,GTW2001,Warren2007} and can be studied by the techniques of 
random matrix theory. The finite temperature version is more difficult to treat, 
but still has nice algebraic properties. In particular the connection to the 
quantum Toda lattice was discovered in \cite{OConnell2012}, which was 
further generalized to the Macdonald process \cite{BC2014}. A few other 
algebraic properties have been discussed in \cite{BCR2013,IS2016}.  

The original OY model is defined for the case where the polymer starts and 
ends at specified positions (point-to-point geometry). One can also consider 
other geometries such as the point-to-line geometry. 
In this paper we consider the model in the stationary situation \cite{OY2001,SeVa2010}.

In our previous paper \cite{IS2017p} we studied the stationary $q$-TASEP. 
We first showed that the $q$-TASEP with a random initial condition can be 
encoded as a marginal of a two-sided version of the $q$-Whittaker process. 
Then by rewriting the Cauchy identity for the ordinary $q$-Whittaker function
and applying the Ramanujan's summation formula and the Cauchy determinant 
identity for the theta function, we were able to find a Fredholm determinant formula 
for the $q$-Laplace transform for a position of the $N$th particle. We also showed 
that the limiting distribution is given by the Baik-Rains distribution $F_{\rm BR}$
\cite{BR2000}. 

In this paper, we discuss the stationary OY polymer model
by taking a scaling limit with $q\rightarrow 1$ of the $q$-TASEP
and the two-sided $q$-Whittaker process. 
We first show that the OY polymer model with boundary sources appears 
as a marginal of a limiting case of the two-sided $q$-Whittaker 
process with two set of parameters. 
We can obtain the stationary OY model by modifying this model with
an appropriate control of the sources as with the 
case of the two-sided $q$-Whittaker process~\cite{IS2017p}.
The same OY model with parameters already appeared in \cite{BCFV2015} but 
the real stationary case was not covered there. We will give a formula for the 
free energy distribution for both the two parameter model and for 
the stationary case.  We also show that the Baik-Rains distribution $F_{\rm BR}$ appears in the long time limit.  

The stationary OY model goes to the stationary KPZ equation under appropriate 
scalings of the model parameters. The latter was already studied in 
\cite{IS2013,BCFV2015} but we will discuss the relations among a few representations. 

The paper is organized as follows. In section 2, we discuss some basic properties of the 
OY model especially focusing on the stationary situation. 
In section 3, we introduce the OY model with boundary sources and state its relation
to the stationary OY model. In section 4, we show that the OY model with boundary sources appears as a limit of a marginal of the two-sided $q$-Whittaker process studied in \cite{IS2017p}. In section 5, we present formulas for the distribution of the free energy 
for the stationary OY model in terms of Fredholm determinant. 
We also study the long time limit and show that the Baik-Rains distribution appears
as the limiting distribution. In section 6, we take a scaling limit to the stationary KPZ 
equation and discuss the connections to previous representations~\cite{IS2013,BCFV2015}.  In Appendix A, we 
summarize basic definitions and properties of the two-sided $q$-Whittaker function 
which are relevant in this paper. In Appendix B, we discuss the inverse Laplace transform. 
Appendix C contains the details of the asymptotic analysis from section 6. 

\smallskip
\noindent
{\bf Acknowledgements.}
The work of T.I. and T.S. is partially supported by JSPS 
KAKENHI Grant Numbers JP25800215, JP16K05192 and JP25103004, JP14510499, 
JP15K05203, JP16H06338 respectively.

\section{The stationary O\rq{}Connell-Yor polymer model}
\label{sOY}
The partition function of the O\rq{}Connell-Yor polymer model is defined by
\begin{align}
Z_{j}(\t)
=
\int_{0<s_1<\cdots<s_{j-1}\le\t}\prod_{i=1}^{j}ds_i
\cdot
e^{\sum_{i=1}^{j}(B_{i}(s_{i})-B_{i}(s_{i-1}))},
\label{l81}
\end{align}
where $s_0=0$, $j\in\Z_+$ and $B_i(\t),~i\in\Z_+$ are the independent standard Brownian motions without drift
\cite{OY2001}.
This can be understood as a partition function of a directed polymer in a random environment 
described by independent Brownian motions,  
which starts at the the site $j=1$ at $\t=0$ and ends at the site $j$ at time $\tau$ 
(point-to-point geometry). 
By using It{\^o}\rq{}s formula, we find that 
it satisfies the discrete stochastic heat equations,
\begin{align}
dZ_{j}(\t)=Z_{j-1}(\t)d\t+Z_{j}(\t)dB_j(\t),
\label{l130d}
\end{align}
where we  interpret the second term as It{\^o} type.
One can extend the values of the index $j$ to the whole $j\in\Z$
and consider the process for  $r_j(\t):=\log Z_{j+1}(\t)-\log Z_{j}(\t)$. 
This process has a stationary measure labeled by a parameter $\a\in\R$
in which all $r_j$'s are independent random variables and each $e^{-r_j}$ 
obeys the Gamma distribution with parameter $\a(>0)$ i.e. the pdf of $e^{-r_j}$ is 
\begin{align}
\P[e^{-r_j}\in dx]=\frac{x^{\a-1}e^{-x}}
{\Gamma(\a)}1_{x>0} dx,
\label{l80}
\end{align} 
see \cite{Spohn2012p}.
We sometimes write~\eqref{l80} as $r_j\sim -\log\Gamma(\a)$. 

Using a version of the Burke's theorem \cite{Burke1956,OY2001,SeVa2010},
one can replace the effects of the whole $Z_j(\t),j \leq 0$
by $Z_1(\t)$ driven by the Brownian motion with drift $\a$. 
This situation with the normalization condition $Z_1(0)=1$ is described by
the SDEs~\eqref{l130d} with $j \geq 1$ 
and $B_1(\t)$ replaced by a standard Brownian motion with drift $\a$.
Let us denote the partition function as $Z_j(\t,\a)$ specifying the
dependence on $\a$.
In~\cite{OY2001,SeVa2010}, it has been shown that it
can be represented as
\begin{align}
Z_j(\t,\a)
=
\int_{-\infty<s_1<\cdots<s_{j-1}\le\t}
e^{\sum_{m=1}^{j}(\tilde{B}_{m}(s_{m})-\tilde{B}_{m}(s_{m-1}))}
\prod_{k=1}^{j-1}ds_k,
\label{l122}
\end{align}
where $s_0=0,~s_j=\t$, and $\tilde{B}_j(s),~j=1,\cdots, N$ are independent two-sided Brownian motions among which $\tilde{B}_1(s)$ has 
drift $\a$ while $\tilde{B}_{j}(s),~j=2,\cdots,N$ have no drifts.
Here the two-sided Brownian motion $\tilde{B}(x)$ with
drift $v$ is defined as
\begin{align}
\tilde{B}(x)
=
\begin{cases}
B_+(x)+vx,& x\ge 0,\\
B_-(-x)+v x, & x<0
\end{cases}
\label{l131}
\end{align}
with $B_{\pm}(x)$ are the independent standard Brownian motions.
Hereafter we call $Z_j(\t,\a)$~\eqref{l122}
the partition function of the stationary O\rq{}Connell-Yor model.

Note that, by the conditioning on the smallest positive $s_k$, rhs of~\eqref{l122} with $j=N$
is rewritten as
\begin{align}
Z_N(\t,\a)=
\sum_{k=1}^N
&
\int_{-\infty<s_1<\cdots<s_{k-1}\le 0}
e^{\sum_{m=1}^{k-1}(B_{m}(s_{m})-B_{m-1}(s_{m-1}))}
\prod_{j=1}^{k-1}ds_j
\notag
\\
&
\times
\int_{0<s_{k}<\cdots<s_{N-1}\le \t}
e^{\sum_{m=k}^{N-1}(B_{m}(s_{m})-B_{m}(s_{m-1}))}
\prod_{j=k}^{N-1}ds_j.
\label{l140}
\end{align}
Since the first factor corresponds to the case $\t=0$ in~\eqref{l122},
it is equal to $e^{\sum_{j=1}^{k-1}y_j}$ in distribution where 
$y_j,~j=1,2,\cdots$ are i.i.d. random variables with $y_j\sim -\log\Gamma(\a)$'s 
while the second one is equal to the partition function $Z_{N-k+1}(\t)$ for the 
point-to-point polymer in (\ref{l81}).
To summarize, we have seen that the partition function of the stationary OY polymer 
with parameter $\a$ can be written as 
\begin{align}
Z_N(\t,\a)
=
\sum_{k=1}^{N} e^{\sum_{j=1}^{k-1} y_{j}}Z_{N-k+1}(\t)
\label{l141}
\end{align}
in distribution,
where the random variables $y_j,~j=1,\cdots,N$ are independent and identically
distributed as $-\log\Gamma (\a)$.

\section{The O\rq{}Connell-Yor polymer model with boundary sources}
\label{OYbs}
Here we introduce a directed random polymer model related to~\eqref{l141},
which has a direct connection to the Whittaker process.
This model is defined as a composition of the OY model with point-to-point geometry 
(with drifts) 
and the log-Gamma discrete random polymer model \cite{Se2012}.
Let us consider a slight modification of (\ref{l81}), in which the polymer starts at site $j=n$
and ends at $j=N$ and the Brownian motions $B_j (t),~j=1,2,\cdots,N$ are the 
independent standard Brownian motions with drift $a_j\in\R$ starting at the origin. 
The partition function of the OY model for this situation is given by 
\begin{align}
Z^{\text{OY}}_{n,N}(\t,a)
=
\int_{0<s_1<\cdots<s_{N-n}\le\t}\prod_{j=1}^{N-n}ds_j
\cdot
e^{\sum_{m=n}^{N}(B_{m}(s_{m-n+1})-B_{m}(s_{m-n}))},
\label{l81d}
\end{align}
where $s_0=0$ (i.e. $B_{n-1}(s_0)=0$) and $s_{N-n+1}=\t$. Note that $Z^{\text{OY}}_{1,N}(\t,\a)=Z_{N}(\t)$ in~\eqref{l81}.

To introduce the log-Gamma discrete random polymer model,  let us 
consider the two dimensional lattice $(i,j),~i=1,\cdots, N, j=1,\cdots,n$.
Let discrete up/right path from $(1,1)$ to $(N, n)$ be an ordered set 
$((i_1,j_1),(i_2,j_2)\cdots, (i_{N+n-1},j_{N+n-1}))$ with $(i_1,j_1)=(1,1)$ and
$(i_{N+n-1},j_{N+n-1})=(N,n)$ such that $(i_k,j_k)\in\Z^2$ and $(i_{k+1}-i_k,j_{k+1}-j_k)\in \{(1,0),(0,1)\}$. The partition function of the log-Gamma polymer model
is defined as
\begin{align}
Z^{\Gamma}_{n,N}(\a,a)=
\sum_{((i_1,j_1),\cdots, (i_{N+n-1},j_{N+n-1}))\in\Omega_{N,n}}e^{\sum_{k=1}^{N+n-1}\o_{i_k,j_k}},
\label{l2}
\end{align}
where $\O_{N,n}$ represents a set of the discrete up/right paths from $(1,1)$ to $(N,n)$ and $\a=(\a_1,\cdots,\a_N)\in\R^N,~a=(a_1,\cdots, a_N)\in\R^{N}$  
in lhs are parameters such that $\a_i-a_j>0$ for 
$i,j=1,\cdots,N$. $\o_{i,j},~i,j=1,\cdots, N$ in rhs
are i.i.d. random variables with $\o_{i,j}\sim-\log\Gamma(\a_i-a_j)$.

\begin{figure}[t]
\begin{center}
\includegraphics[scale=0.6]{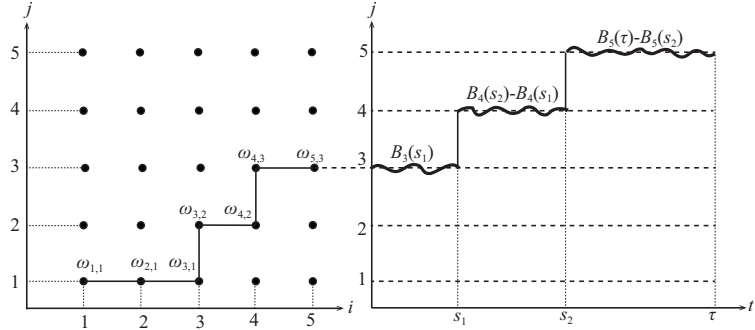}
\caption{\label{figOYbs}
The O'Connell-Yor polymer model with boundary sources. An example with $N=5,n=3$. 
}
\end{center}
\end{figure}

In terms of the two polymers above, a semi-discrete polymer model 
is defined as follows. 
\begin{definition}(\cite{BCFV2015})
The partition function of the O'Connell-Yor polymer model
with boundary sources is defined as
\begin{align}
Z(\t,\a,a)=\sum_{n=1}^NZ^{\Gamma}_{n,N}(\a,a)Z^{\rm{OY}}_{n,N}(\t,a).
\label{l1}
\end{align}
\end{definition}
\noindent
The first part $Z^{\Gamma}_{n,N}(\a,a)$ can be regarded as representing boundary sources. 
See Fig.  \ref{figOYbs}.

\noindent
When we set $\a_1,\cdots,\a_N\rightarrow\i$ and $a_1=\cdots=a_N=0$, we see from~\eqref{l80} 
that the whole weights $e^{\o_{i,j}}$\rq{}s in the log-Gamma polymer model~\eqref{l2} vanish. 
Noting that in~\eqref{l2}, the number of lattice points, to which we assign the weights
is $N+n-1$, which increases with $n$, we find that in~\eqref{l1}, the contribution of $n=1$ becomes dominant. (In other word, in Fig.~\ref{figOYbs}, the path crossing the bottom
points $(1,1),~(2,1),\cdots,(5,1)$ in the left plane becomes dominant.)
Thus in this limit $Z(\t,\a,a)$ reduces to the OY model without sources,
$Z_{1,N}^{\text{OY}}(\t,0)$.

This model is related to the stationary OY model~\eqref{l122} in the following 
way. To describe the stationary situation we need to specialize 
the parameters of the OY model with boundaries sources as 
\begin{align}
\a_1=\a, \a_2, \cdots, \a_N\rightarrow\infty,~a_1=a,a_2=\cdots=a_N=0
\label{l42}
\end{align}
and take the limit $a\rightarrow \a$.
However note that in this limit, the model is not well-defined 
since $\o_{1,1}\sim \log -\Gamma(\a-a)$ becomes singular in the limit. Thus 
we introduce the modified model which is defined in a same way as
the original one~\eqref{l1} except that $\o_{1,1}=0$. We write the
partition function of the modified model as $Z^{(0)}(\t,\a,a)$. Note that 
it is related to $Z(\t,\a,a)$~\eqref{l1} as
$
Z(\t,\a,a)=e^{\o_{1,1}}Z^{(0)}(\t,\a,a).
$
Considering~\eqref{l141},
we find under~\eqref{l42}
\begin{align}
\lim_{a\rightarrow\a}Z^{(0)}(\t,\a,a)= Z_N(\t,\a)
\label{l133}
\end{align}
in distribution.
In this way we can study the stationary OY model by considering a limiting case 
of the OY model with boundary sources.

\section{OY model with boundary sources and the Whittaker process}
In \cite{IS2017p}, we studied the stationary $q$-TASEP using a two-sided 
version of the $q$-Whittaker process. In this section we will see that the
$q$-Whittaker functions with signatures (see Definitions~\ref{d1} and~\eqref{d2}) 
go to the Whittaker functions with two sets of 
parameters, which is previously shown to be related to the OY polymer 
with the boundary sources in \cite{BCFV2015}. This opens the way to 
study the stationary OY model by considering a limit of the analysis 
in \cite{IS2017p} using the two-sided $q$-Whittaker process. 
In Appendix A, we give a brief summary of the definitions and properties of 
the two-sided $q$-Whittaker functions and process, which are used in this paper.  

The  Whittaker process  is defined as follows.
Let $Y\in\R^{N(N+1)/2}$ be a triangular array $Y=(y^{(1)},\cdots, y^{(N)})$
where $y^{(k)}=(y^{(k)}_1,\cdots,y^{(k)}_k)$ with $y^{(j)}_i\in\R$ for
$1\le i\le j\le N$. 
The Whittaker function $\tilde{\Psi}_{\nu}(y^{(N)})$ with parameter 
$\nu=(\nu_1,\dots,\nu_N)\in\R^N$ has the following integral representation~\cite{Givental1997},
\begin{align}
\Psi_{\nu}(y^{(N)})
&=
\int_{\R^{N(N-1)}}e^{\mathcal{F}_{\nu}(Y)} \prod_{1\le j\le k\le N-1}dy^{(k)}_j,
\label{l82}
\end{align}
where $\mathcal{F}_{\nu}(Y)$ 
is defined by
\begin{align}
\mathcal{F}_{\nu}(Y)
&=
i\sum_{k=1}^N\nu_k
\left(
\sum_{j=1}^k y^{(k)}_j
-
\sum_{j=1}^{k-1}
y^{(k-1)}_j
\right)
-
\sum_{k=1}^{N-1}
\sum_{j=1}^k
\left(
e^{y^{(k)}_j-y^{(k+1)}_j}+e^{y^{(k+1)}_{j+1}-y^{(k)}_j}
\right).
\label{l83}
\end{align}
By definition, one sees that $\Psi_{\nu}(y^{(N)})$ is symmetric 
in $\nu=(\nu_1,\cdots,\nu_N)$.

We also define the function $\theta_{\mu,\t}(y^{(N)})$ with
parameters $\mu=(\mu_1,\cdots,\mu_N)\in \R^{N}$
and $\t>0$,
\begin{align}
\theta_{\mu,\t}(y^{(N)})
&=
\int_{\R^N}
\prod_{j=1}^N d\nu_j\cdot
\Psi_{\nu}(y^{(N)})e^{-\t\sum_{j=1}^N\nu_j^2/2}
\cdot
\prod_{m,n=1}^N
\Gamma (\mu_m+i\nu_n)
\cdot
m_N(\nu),
\label{l84}
\end{align}
where the Sklyanin measure $m_N(\nu)$ is defined by
\begin{align}
m_N(\nu)&=\frac{1}{(2\pi)^N}\prod_{i\neq k}\frac{1}{\Gamma (i\nu_k-i\nu_j)}.
\label{l85}
\end{align}
As will be shown below in the proof of Proposition~\ref{lpp2}, the functions
$\Psi_{\nu}(y^{(N)})$~\eqref{l82} and $\theta_{\mu,\t}(y^{(N)})$~\eqref{l84}
can be regarded as the $q\rightarrow 1$ scaling limit of $P_{\l}(a)$~\eqref{Wh}
and $Q_{\l}(\a,t)$~\eqref{defQ} respectively (see~\eqref{lpp22} and~\eqref{lpp24} below.)

The Whittaker process with parameters $a,\a\in\R^N$ and $\t>0$
is defined in terms of~\eqref{l82} and~\eqref{l84} 
as follows~\cite{BCFV2015}:
\begin{definition} 
For $a,\a\in\R^N$ such that $a_i+\a_j>0$ for $1\le i,j\le N$ and $\t>0$,
the Whittaker process is defined as a probability measure
on $\R^{N(N+1)/2}$ with the pdf
\begin{align}
W_{a;\a,\t}(Y)
=
e^{\mathcal{F}_{i a}(Y)}
\theta_{\a,\t}(y^{(N)})
e^{-\t\sum_{j=1}^Na_j^2/2}
\prod_{m,n=1}^N
\frac{1}{\Gamma(\a_m+a_n)}.
\label{l86}
\end{align}
\end{definition}
\smallskip
\noindent
In the limit as $\a_j\rightarrow\i,~j=1,\cdots,N$, the density function~\eqref{l86} 
reduces to the one in~\cite{OConnell2012, BC2014},
\begin{align}
e^{\mathcal{F}_{i a}(Y)}
\int_{\R^N}
\prod_{j=1}^N d\nu_j\cdot
\Psi_{-\nu}(y^{(N)})e^{-\t\sum_{j=1}^N\nu_j^2/2}
m_N(\nu)
\cdot
e^{-\t\sum_{j=1}^Na_j^2/2}.
\label{l109}
\end{align}

Furthermore from~\eqref{l82} and~\eqref{l86}, we immediately have the following
\begin{proposition} 
\label{lp7}
The pdf of the marginal density of $W_{\a;a,\t}(Y)$~\eqref{l86} on
$y^{(N)}$ is expressed as
\begin{align}
\Psi_{i a}(y^{(N)})
\theta_{\a,\t}(y^{(N)})
e^{-\t\sum_{j=1}^Na_j^2/2}
\prod_{m,n=1}^N
\frac{1}{\Gamma(\a_m+a_n)}.
\label{lp71}
\end{align}
\end{proposition}

\smallskip
\noindent
We call~\eqref{lp7} the Whittaker measure.

We will show that the Whittaker process $W_{-a;\a,\t}(Y)$ appears as a limit of our 
two-sided $q$-Whittaker process~\eqref{2qWh-pr}
and one can study the OY model with boundary sources~\eqref{l1} by considering the limit 
of our results for the $q$-TASEP in \cite{IS2017p}. 
In this section, we rewrite the parameters $a_j$ and 
$\a_j,~j=1,\cdots,N$ in~\eqref{2qWh-pr} 
as $\tilde{a}_j$ and $\tilde{\a}_j$ to distinguish them from $a_j,~\a_j$
in $W_{-a;\a,\t}(Y)$. We scale each variable and parameter of~\eqref{2qWh-pr} 
as 
\begin{align}
&q=e^{-\e},~t=\t \e^{-2},~\tilde{a}_j=e^{-\e a_j},~\tilde{\a}_j=e^{-\e\a_j},
\notag
\\
&\l^{(k)}_j=\t\e^{-2}+(k+1-2j-N)\e^{-1}\log\e^{-1}+y^{(k)}_j\e^{-1},
\label{l4}
\end{align}
for $j=1,2,\cdots,N$ and taking the limit $\e\rightarrow 0$.
Here we assumed $0<\a_j<\infty, 1\leq j\leq N$. When we study the case 
where only $M$ of them are finite and $\a_j\to\infty, N-M+1\leq j\leq N$, 
we should set $\tilde{\a}_j=0, N-M+1\leq j\leq N$ and replace 
$-N\e^{-1}\log\e^{-1}$ by $-M\e^{-1}\log\e^{-1}$. 
This is an aspect 
which is different from the previously studied scaling limit from 
$q$-Whittaker  processes to the Whittaker process \cite{BC2014,BCFV2015}. 
There the term $-M\e^{-1}\log\e^{-1}$ is replaced by $+M\e^{-1}\log\e^{-1}$.
The minus sign of $-M\e^{-1}\log\e^{-1}$ results from
the ``two-sided'' nature of~\eqref{2qWh-pr}: our process is defined 
on the signature $\mathcal{S}_n$~\eqref{signature} of which
each element can take negative value. Due to this property, 
the scaling changes to the minus direction. 

We obtain the following
\begin{proposition}
\label{lpp2}
Under the scaling~\eqref{l4}, 
the $\e\rightarrow 0$ limit  of $P_t(\underline{\l}_N)$~\eqref{2qWh-pr}
becomes
\begin{align}
\lim_{\e\rightarrow\i}\e^{\frac{N(N+1)}{2}} P_t(\underline{\l}_N)
=W_{-a;\a,\t}(-Y),
\label{lpp21}
\end{align}
where $-Y=(-y^{(1)},\cdots,-y^{(N)})$ with $-y^{(j)}=(-y^{(j)}_j,\cdots,-y^{(j)}_1)$,
$j=1,\cdots,N$.
\end{proposition}

\smallskip
\noindent
{\bf Proof.} 
For comparing~\eqref{l82} with~\eqref{Wh},
we write $e^{\mathcal{F}_{\nu}(Y)}$ in~\eqref{l82}
as
\begin{align}
e^{\mathcal{F}_\nu(Y)}=\prod_{j=1}^N\Psi_{\nu_j}(y^{(j-1)},y^{(j)}),
\label{lpp215}
\end{align}
where
\begin{align}
\Psi_{\nu_j}(y^{(j-1)},y^{(j)})
=
\exp\left(
i\nu_j
\left(
\sum_{i=1}^j y^{(j)}_i
-
\sum_{i=1}^{j-1}
y^{(j-1)}_i
\right)
-
\sum_{i=1}^{j-1}
\left(
e^{y^{(j-1)}_i-y^{(j)}_i}+e^{y^{(j)}_{i+1}-y^{(j-1)}_i}
\right)
\right).
\label{lpp23}
\end{align}
We will show that the skew $q$-Whittaker function $P_{\l^{(j)}/\l^{(j-1)}}(\tilde{a}_j)$
with some factors goes to this function~\eqref{lpp23}, i.e.
\begin{align}
&\lim_{\e\rightarrow 0}
\frac{e^{(j-1)\mathcal{C}(\e)}}{\tilde{a}_j^{t+N\e^{-1}\log \e}}
P_{\l^{(j)}/\l^{(j-1)}}(\tilde{a}_j)
=
\Psi_{i a_j}(y^{(j-1)},y^{(j)})
=
\Psi_{-i a_j}(-y^{(j-1)},-y^{(j)}),
\label{lpp22}
\end{align}
where
$
\mathcal{C}(\e)=
-\frac{\pi^2}{6}\e^{-1}-\frac12 \log \frac{\e}{2\pi}.
$
Furthermore we will also show that 
\begin{align}
&\lim_{\e\rightarrow 0}
e^{\frac{N(N+1)}{2}\mathcal{C}(\e)-Nt}\e^{\frac{N(N+1)}{2}}
\prod_{j=1}^N
\tilde{\a}_j^{N\e^{-1}\log\e}
\cdot
Q_{\l^{(N)}}(\tilde{\a},t)
=
\theta_{\a,\t}(-y^{(N)}),
\label{lpp24}
\\
&\lim_{\e\rightarrow 0}
e^{N t-N^2\mathcal{C}(\e)}
\prod_{j=1}^N
\frac
{\tilde{a}_j^{t+N\e^{-1}\log\e}}
{\tilde{\a}_j^{N\e^{-1}\log\e}}
\cdot
\frac
{1}
{\Pi (\tilde{a};\tilde{\a},t)}
=
\frac
{1}
{\prod_{j=1}^Ne^{\t a_j^2/2}\prod_{k=1}^N\Gamma(\a_k-a_j)}.
\label{lpp25}
\end{align}
Then (\ref{lpp21}) immediately follows from~\eqref{lpp22}--\eqref{lpp25}. 

Hereafter we give proofs of~\eqref{lpp22}--\eqref{lpp25}. Limiting behaviors 
of various factors can be taken from \cite{BC2014}. 

\smallskip
\noindent
{\it Proof of~\eqref{lpp22}.}
Here we show the first equality since
the second equality follows immediately by definitions of $\Psi_{\nu}(y^{(N)})$~\eqref{l82}
and $-y^{(N)}$ written below~\eqref{lpp21}. 
Substituting~\eqref{l4} into~\eqref{skewWh} with $\l=\l^{(j)}$, $\mu=\l^{(j-1)}$,
and $a=\tilde{a}_j$,
we have
\begin{align}
P_{\l^{(j)}/\l^{(j-1)}}(\tilde{a}_j)
=&
\tilde{a}_j^{t+N\e^{-1}\log\e}
\prod_{i=1}^{j-1}
\frac
{(q;q)_{-2\e^{-1}\log\e+\e^{-1}(y^{(j)}_i-y^{(j)}_{i-1})}}
{(q;q)_{-\e^{-1}\log\e+\e^{-1}(y^{(j)}_i-y^{(j-1)}_{i})}
 (q;q)_{-\e^{-1}\log\e+\e^{-1}(y^{(j-1)}_i-y^{(j)}_{i+1})}
}
\notag
\\
&\times
\exp\left(
-a_j
\left(
\sum_{i=1}^j y^{(j)}_i
-
\sum_{i=1}^{j-1}
y^{(j-1)}_i
\right)\right).
\label{lpp27}
\end{align}
Here we see, for $c>0$ and $y\in\R$
\begin{align}
(q;q)_{-c\e^{-1}\log\e+\e^{-1}y}
=
e^{\mathcal{C}(\e)+\e^{c-1}e^{-y}}
\label{lpp28}
\end{align}
by Corollary 4.10 in~\cite{BC2014}.  Applying this to the second factor in~\eqref{lpp27},
we get~\eqref{lpp22}.

\smallskip
\noindent
{\it Proofs of~\eqref{lpp24} and~\eqref{lpp25}.}
For showing~\eqref{lpp24}, we consider the limiting behavior
of each factor in the definition of $Q_{\l^{(N)}}(\tilde{\a};t)$~\eqref{defQ}.
Hereafter we change integration variables $z_j$ in~\eqref{defQ} to
$z_j=e^{iw_j},~j=1,\dots,N$.
First, one sees that under the scaling~\eqref{l4},
\begin{align}
\lim_{\e\rightarrow0}
\prod_{i=1}^{N-1}(q^{\l^{(N)}_i-\l^{(N)}_{i+1}+1};q)_{\infty}=1
\label{lpp29}
\end{align}
from Lemma 4.25 in~\cite{BC2014}. 
Next for $P_{\l^{(N)}}(1/z)$ we use~\eqref{lpp22} and have
\begin{align}
\lim_{\e\rightarrow\i}
\e^{\frac{N(N-1)}{2}}e^{\frac{N(N-1)}{2}\mathcal{C}(\e)}
\prod_{j=1}^N z_j^{t+N\e^{-1}\log \e}
\cdot
P_{\l^{(N)}}(1/z)
=\Psi_{-w}(y^{(N)})=\Psi_{w}(-y^{(N)}).
\label{lpp210}
\end{align}
For $\Pi(z;\tilde{\a},t)$~\eqref{d14}, we have
\begin{align}
\Pi(z;\tilde{\a},t)
=
\prod_{i,j=1}^N
\frac
{1}
{(e^{-\e (\a_i+i w_j)};e^{-\e})_{\infty}}
\cdot
\prod_{j=1}^N
e^{e^{i\e w_j}\e^{-2}\t}.
\label{lpp211}
\end{align}
Using the relations
\begin{align}
\lim_{\e\rightarrow 0}
\frac{e^{\mathcal{C}(\e)}\e^{1-x}}{(e^{-\e x};e^{-\e})_{\infty}}
=\Gamma (x),
~~
\lim_{\e\rightarrow 0}
\left(e^{i\e w_j}\e^{-2}\t-\e^{-2}\t-i\e^{-1}w_j\t\right)=-\t w_j^2/2,
\label{lpp212}
\end{align}
where the first one is given in (4.55) in~\cite{BC2014},
we have
\begin{align}
\lim_{\e\rightarrow 0}
e^{N^2\mathcal{C}(\e)-Nt}
\e^{N^2}
\prod_{i,j=1}^N
\left(\frac{\tilde{\a_i}}{z_j}
\right)^{\e^{-1}\log\e}
\cdot
\prod_{j=1}^N
\frac
{1}
{z_j^t}
\cdot
\Pi (z;\tilde{\a},t)
=
\prod_{i,j=1}^N
\Gamma(\a_i+iw_j)
\cdot
\prod_{j=1}^N
e^{-\t w_j^2/2}.
\label{lpp213}
\end{align}
At last for $m_N^q(z)\prod_{j=1}^Ndz_j/z_j$, we find 
\begin{align}
\lim_{\e\rightarrow 0}
\e^{-N^2}e^{-N(N-1)\mathcal{C(\e)}}
m_N^q(z)
\prod_{j=1}^N
\frac{dz_j}{z_j}
=
m_N(w)
\prod_{j=1}^N dw_j
\label{lpp214}
\end{align}
by using (4.36) in~\cite{BC2014}.
Combining~\eqref{d14} with the scaling limits~\eqref{lpp29},~\eqref{lpp210},~\eqref{lpp212}, and~\eqref{lpp214}, 
we arrive at~\eqref{lpp24}.
Then \eqref{lpp25} can be obtained by~\eqref{lpp213} with $z$ and $iw_j$ replaced 
by $1/\tilde{a}$ and $-a_j$ respectively.
\qed

Thus we have shown that the Whittaker process with two parameters appears as a 
limit of our two-sided $q$-Whittaker process. 
In addition the relationships between~\eqref{l1} and~\eqref{l86} is also known:
\begin{proposition} 
\label{lpp1}
$\log Z(\t,\a,a)$ has the same distribution as 
$y^{(N)}_1$ in the Whittaker process $W_{-a;\a,\t}(Y)$~\eqref{l86}.
\end{proposition}

\smallskip
\noindent
This relation was obtained by introducing a version of $q$-Whittaker process, 
which is different from ours~\eqref{2qWh-pr} and by taking $q\rightarrow 1$ scaling limit
\cite{BCFV2015}. 

From Propositions~\ref{lpp2} and~\ref{lpp1}, we find
a relation between $\l^{(N)}_N$ in~\eqref{det_formula} and
the OY polymer with the boundary sources $Z(\t,\a,a)$~\eqref{l1}.
From the definition of $-Y$ (see below~\eqref{lpp21}), 
one sees that the marginal density of 
$W_{-a,\a,\t}(Y)$ on $y^{(N)}_1$ is equal to that of 
$W_{-a,\a,\t}(-Y)$ on $-y^{(N)}_N$. 
Combining this 
with Proposition~\ref{lpp2}, we conclude that the marginal density function of 
$W_{-a,\a,\t}(Y)$ on $-y_1^{(N)}$ 
is given by  the scaling limit~\eqref{l4} of the marginal density of 
$P_t(\underline{\l}_N)$ on $\l^{(N)}_N$, 
which describes the $N$th particle of the $q$-TASEP.
Furthermore combining this with the proposition \ref{lpp1} 
we see that the marginal density of $P_t(\underline{\l}_N)$ on $\l^{(N)}_N$
(or equivalently the marginal density of the two-sided $q$-Whittaker 
measure~\eqref{2qWh-me} on $\l_N$)
goes to the density function of $-\log Z(\t,\a,a)$
under the scaling limit~\eqref{l4} with $\e\rightarrow 0$.
Thus in this limit, Theorem~\ref{ql-det} becomes a relation
on $Z(\t,\a,a)$. We have the following:

\begin{proposition}
\label{pp3}
The Laplace transform of $Z(\t,\a,a)$ (\ref{l1}) is written as the Fredholm determinant, 
\begin{align}
\left\langle e^{-u Z(\t,\a,a)}\right\rangle
=
\det
\left(
1-f_u K
\right)_{L^2(\R)}.
\label{l5}
\end{align}
Here in lhs, $u\in\C$ with $\text{Re~} u>0$, $\langle\cdot\rangle$ 
represents the average over
the random variables $\o_{ij}, 1\le i,j\le N$ and $B_i,~i=1,\cdots,N$
in~\eqref{l1} and on rhs the kernel $f_u K$ is given by 
\begin{align}
&f_u(x)= \frac{1}{1+e^{-x}/u}, 
\label{l6}
\\
&K(x_1,x_2)
=
\sum_{l=0}^{N-1}\phi_l(x_1)\psi_l(x_2),
\label{l7}
\end{align}
where the functions $\Phi_l(x)$ and $\Psi_l(x)$ are given as
\begin{align}
&\phi_l(x)
=
\frac{1}{2\pi i}\oint dv
\frac
{e^{v x-v^2\t/2}}
{v-a_{l+1}}
\prod_{i=1}^l
\frac
{v-\a_i}
{v-a_i}
\prod_{j=1}^N
\frac
{\Gamma(1+v-a_j)}
{\Gamma(1+\a_j-v)},
\label{l8}
\\
&\psi_l(x)
=
\frac{(\a_{l+1}-a_{l+1})}{2\pi }
\int_{-\infty}^{\infty}dw
\frac
{e^{-iw x-w^2\t/2}}
{\a_{l+1}-iw}
\prod_{k=1}^l
\frac
{iw-a_k}
{iw-\a_k}
\prod_{j=1}^N
\frac
{\Gamma(1+\a_j-i w)}
{\Gamma(1+i w -a_j)},
\label{l9}
\end{align}
where in~\eqref{l8}, the contour encloses $a_j,~j=1,\cdots,N$ positively.
\end{proposition}
\smallskip
\noindent
In the proof below, we take the $q\rightarrow 1$ scaling limit 
of~\eqref{det_formula}. As with $\tilde{a}_j$ and $\tilde{a}_j$ in~\eqref{l4},
we rewrite $\phi_l(n)$~\eqref{phi} and $\psi_l(n)$~\eqref{psi} as 
$\tilde{\phi}_l(n)$ and $\tilde{\psi}_l(n)$ to distinguish them from~\eqref{l8}
and~\eqref{l9} respectively.

\smallskip
\noindent
{\bf Proof.}
We consider the $\e\rightarrow 0$ limit of~\eqref{det_formula} 
under the scaling~\eqref{l4} and
\begin{align}
\zeta=-\e^{2N}ue^{\t\e^{-1}}. 
\label{pp31}
\end{align}

First substituting~\eqref{l4} and~\eqref{pp31} into lhs of~\eqref{det_formula}
we have
\begin{align}
\lim_{\e\rightarrow 0}\frac{1}{(\zeta q^{\l_N};q)_{\infty}}
=
\lim_{\e\rightarrow 0}
e_q\left(x_q\right)|_{q=e^{-\e}}
=e^{-u e^{-y^{(N)}_N}},
\label{pp32}
\end{align}
where $e_q(x_q)=1/((1-q)x_q;q)_\i$ is the $q$-exponential function with 
$x_q=-\e u e^{-y^{(N)}_N}/(1-q)$ 
and we used the fact $\lim_{q\rightarrow 1}e_{q}(x)=e^x$
uniformly on $x\in(-\infty,0)$.
Thus from the remark below Proposition~\ref{lpp1}, we have
\begin{align}
\lim_{\e\rightarrow 0}
\left\langle
\frac{1}{(\zeta q^{\l_N};q)_{\infty}}
\right\rangle
=
\left\langle e^{-u Z(\t,\a,a)}\right\rangle
\label{l90}
\end{align}
under the scalings~\eqref{l4} and~\eqref{pp31}.

Next we consider rhs of~\eqref{det_formula}.  We begin with the function
$f(n)$~\eqref{f}. Associated with~\eqref{pp31},  
we scale $n$ as
\begin{align}
n=-\t\e^{-2}-2N\e^{-1}\log \e+x \e^{-1}.
\label{pp33}
\end{align}
Substituting $q=e^{-\e}$,~\eqref{pp31} and~\eqref{pp33} into~\eqref{f}, one immediately sees $\lim_{\e\rightarrow 0}f(n)=f_u(x)$.

Next we show that under~\eqref{l4} and~\eqref{pp33}
\begin{align}
\lim_{\e\rightarrow 0}
e^{\t \e^{-2}}\prod_{j=1}^N\e^{\a_j+a_j}\cdot
\tilde{\phi}_l(n)=\phi_l(x),~
\lim_{\e\rightarrow 0}
e^{-\t\e^{-2}}\prod_{j=1}^N\frac{1}{\e^{\a_j+a_j}}\cdot
\tilde{\psi}_l(n)=\psi_l(x)
\label{pp35}
\end{align}
by simple saddle point analyses.
Here we consider only the case of $\tilde{\phi}_l(n)$ since that of $\tilde{\psi}_l(n)$ 
can be obtained in a similar way. Substituting the scalings~\eqref{l4} 
and~\eqref{pp33} into the definition of $\tilde{\phi}_{l}(n)$~\eqref{phi}, we have
\begin{align}
\tilde{\phi}_{l}(n)
&=
\int_D
\frac{dv}{2\pi i}
\frac
{e^{-\frac{\t}{\e^2}g(v)}}
{v^{\e^{-1}x-2N\e^{-1}\log \e+N}}
\frac
{1}
{v-q^{a_{l+1}}}
\prod_{j=1}^l
\frac
{v-q^{\a_j}}
{v-q^{a_j}}
\notag\\
&~~~~\times
\prod_{j=1}^N
\frac
{\Gamma_q(1-a_j+\log_q v)}
{\Gamma_q(1-\a_j-\log_qv)}
(1-q)^{2w-\a_j-a_j},
\label{l10}
\end{align}
where $g(v)=v-\log v$ and we used the $q$-Gamma function 
$\Gamma_q(x)=
(1-q)^{1-x}
\frac
{(q;q)_{\infty}}
{(q^x;q)_{\i}}$. 
Noting the saddle point $v_c$ such that $g\rq{}(v_c)=0$ is 1, we scale $v$
around the saddle point,
\begin{align}
v=q^{w}=e^{-\e w}.
\label{l15}
\end{align}
Thus we find
\begin{align}
&\lim_{\e\rightarrow 0}-\frac{\t}{\e^2}(g(v)-1)=\lim_{\e\rightarrow 0}\frac{-\t g\rq{}\rq{}(1)}{2\e^2}(v-1)^2+O(\e)=-\frac{\t w^2}{2},
\label{l12}
\\
&\lim_{\e\rightarrow 0}\frac{1}{v^{\e^{-1}x}}=e^{w x},
\label{l13}
\\
&
\lim_{\e\rightarrow 0}
\frac
{dv}
{v-q^{a_{l+1}}}
\prod_{j=1}^l
\frac
{v-q^{\a_j}}
{v-q^{a_j}}
=
\frac{dw}{w-a}
\prod_{j=1}^l
\frac
{w-\a_j}
{w-a_j},
\label{l14}
\\
&
\lim_{\e\rightarrow 0}
\frac{1}{v^{-2N\e^{-1}\log\e+N}}
\prod_{j=1}^N(1-q)^{2w-\a_j-a_j}\e^{\a_j+a_j}
=1.
\label{pp34}
\end{align}
Furthermore noting $\lim_{q\rightarrow 1}\Gamma_q(x)=\Gamma(x)$ and
$\log_q v=w$, we have
\begin{align}
\lim_{\e\rightarrow 0}
\prod_{j=1}^N
\frac
{\Gamma_q(1-a_j+\log_q v)}
{\Gamma_q(1-\a_j-\log_qv)}
=
\prod_{j=1}^N
\frac
{\Gamma(1-a_j+w)}
{\Gamma(1-\a_j-w)}.
\label{l16}
\end{align}
From~\eqref{l12}--\eqref{l16}, we arrive at the first relation in~\eqref{pp35}.
\qed
\smallskip

The relation~\eqref{l5} is a generalization of Proposition 12 in~\cite{IS2016} to
the case with two sets of boundary parameters $\a=(a_1,\cdots,a_N),~a=(a_1,\cdots,a_N)$.
In the limiting case $\a_1,\cdots,\a_N\rightarrow\i$, $a_1=\cdots=a_N=0$, one finds that~\eqref{l7} reduces to $K(x,y;\t)$ (4.10) in~\cite{IS2016}.

\section{Distributions in the  stationary O\rq{}Connell-Yor model}
\label{OYsta}
By Proposition~\ref{lpp1}, we have found that the law of $\log Z(\t,\a,a)$ is the same
as the marginal low on $y^{(N)}_1$ of the $q$-Whittaker process~\eqref{l86} with
$a$ replaced by $-a$
or equivalently of the $q$-Whittaker measure~\eqref{lp71}
with $a$ replaced by $-a$. Note that~\eqref{lp71} is symmetric in
$a=(a_1,\cdots,a_N)$ and $\a=(\a_1,\cdots,\a_N)$.
Due to the symmetry 
the specialization~\eqref{l42} is equivalent to 
\begin{align}
\a_1, \cdots, \a_{N-1}\rightarrow\infty,~\a_N=\a,~a_1=\cdots=a_{N-1}=0,~a_N=a.
\label{l121}
\end{align}
Hereafter we adopt~\eqref{l121}.  Note that $a$, $\a$ are real numbers 
rather than the shorthanded notation $a=(a_1,\cdots,a_N),\a=(\a_1,\cdots,\a_N)$
in section
~\ref{OYbs}. For $Z(\t,\a,a)$~\eqref{l1} and $Z^{(0)}(\t,\a,a)$ defined 
below~\eqref{l42},  we define
\begin{gather}
F(y) = \P[\log Z(\tau,\a,a) \leq y],
~~
F_0(y) = \P[\log Z^{(0)}(\tau,\a,a)\leq y],
\label{FF0}
\\
G(u)
=
\left\langle
e^{-uZ(\t,\a,a)}
\right\rangle,
~~
G_0(u)
=
\left\langle
e^{-uZ^{(0)}(\t,\a,a)}
\right\rangle.
\label{GG0}
\end{gather}
Note that the averages on rhs in (\ref{GG0}) are different for $G(u)$ and $G_0(u)$
and they are with respect to the unmodified and the modified model respectively. 
To consider the stationary limit~\eqref{l121} with $a\rightarrow \a$, 
we need to have
the relations which connect $Z(\t,\a,a)$ and the modified 
one $Z^{(0)}(\t,\a,a)$. 
We use the results from Appendix B.2. 
For the OY model, the random variable $\chi$ 
is distributed according to $-\log \Gamma (\nu)$ with parameter $\nu=\a-a$.
(For the definition of $-\log \Gamma (\nu)$, see~\eqref{l80}.)
Its Laplace (or Fourier for $\xi\in i\R$) transform is 
\begin{equation}
 g(\xi) 
 = 
 \langle e^{-\xi \chi} \rangle 
 =
 \int_{\R} dx \frac{x^{\xi+\nu-1}e^{-x}}{\Gamma(\nu)}
 =
 \frac{\Gamma(\nu+\xi)}{\Gamma(\nu)}. 
 \label{gGamma}
\end{equation}

 \smallskip
One can find an expression for $F_0(y)$~\eqref{FF0} in terms of $G(u)$~\eqref{GG0}. 
\begin{proposition}
\label{lp1}
Let $Z(\t,\a,a)$ (resp. $Z^{(0)}(\t,\a,a)$) be the partition function of the OY polymer model 
(\ref{l1}) when the parameters are given by (\ref{l42}), $\a>a$  (resp. and  
$\o_{1,1}$ in (\ref{l2}) is set to be zero). 
The distribution function $F_0(y)$ for $\log Z^{(0)}$ (\ref{FF0}) is recovered from $G(u)$~\eqref{GG0} by 
the following formula. 
\begin{align}
F_0(y)
=
\int_{i\R}\frac{d\xi}{2\pi i}
\frac
{\Gamma(\a-a) e^{y\xi}}
{\Gamma(\a-a+\xi)\Gamma(1+\xi)}
\int_0^\i
u^{\xi-1} G(u) du.
\label{lp13}
\end{align}
\end{proposition}

\smallskip
\noindent
{\bf Remark.} 
A similar formula was obtained in \cite{BCFV2015} for the stationary KPZ equation
using a property of the 0th Bessel function which appears for this special case. 
Here we show that the formula is a consequence of a combination of a few basic facts. 
 
\smallskip
\noindent
{\bf Proof.} 
Let us set the distribution function $F(y),~y\in\R$ 
in the argument below~\eqref{invLap3} to be the one in~\eqref{FF0}.
(In this case $\vp(x)$ in Appendix~\ref{sinvLap} becomes
$\vp(x) := F(\log x)=\P (Z(\t,\a,a)\leq x), x>0$, which is the
distribution function of $Z(\t,\a,a)$.)
By (\ref{invLap2FF}), one has
\begin{equation}
F^{\sharp}(\xi) = \frac{1}{\Gamma(\xi+1)} \int_0^{\i} u^{\xi-1} G(u) du. 
\label{Fhat}
\end{equation}
where $F^{\sharp}(\xi)$ is the Fourier transform of $F(y)$ (see~\eqref{invLap2FF} for 
more detailed argument.)
On the other hand, due to $Z(\t,\a,a)=Z^{(0)}(\t,\a,a) e^{\chi}$, we find 
\begin{equation}
 F_0^{\sharp}(\xi) = \frac{\Gamma(\a-a)}{\Gamma(\a-a+\xi)} F^{\sharp}(\xi).  
 \label{F0hat}
\end{equation}
Combining (\ref{Fhat}),(\ref{F0hat}) and applying the inverse Fourier transform, 
we arrive at (\ref{lp13}). 
\qed

For the case of the OY polymer model, the distribution function is infinitely differentiable
since it is expressed as the marginal distribution of the Whittaker measure~\eqref{lp71}
with $a$ replaced by $-a$ on $y^{(N)}_1$ and the Whittaker function~\eqref{l82} appearing
in~\eqref{lp71} is infinitely differentiable with respect to each $y^{(N)}_j,~j=1,\cdots,N$.
We have
\begin{corollary}
\begin{equation}
 F_0(y) 
 = 
 \Gamma(\nu)
 \sum_{n=0}^{\i} \frac{1}{n!}\frac{d^n}{d\nu^n} \left( \frac{1}{\Gamma(\nu)} \right) F^{(n)}(y),
\label{F0dFd}
\end{equation}
where $F^{(n)}$ means the $n$th derivative of $F(y)$. 
\end{corollary}

\smallskip
\noindent
{\bf Remark.} 
Note that on rhs one can use any representation of $F(y)$. For example, 
if one employs the formula (\ref{invLap3}), there is no complex integral 
and hence~\eqref{F0dFd} with this formula is useful for numerical evaluation.  
Formally (\ref{F0dFd}) can be written as 
\begin{equation}
F_0(y) = \frac{\Gamma(\a-a)}{\Gamma(\a-a+d/dy)}F(y).
\label{F0dF}
\end{equation}
This type of formula was obtained for the stationary KPZ equation in \cite{IS2013}. 
Though it may look awkward with the derivative in the denominator, 
it has a solid meaning and is practically useful as explained above. 

\begin{figure}[t]
\begin{center}
\includegraphics[scale=0.6]{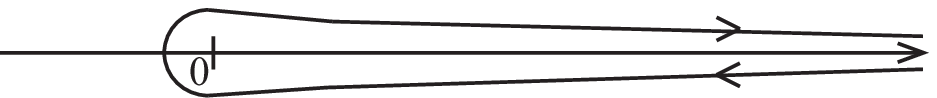}
\caption{\label{figgcontour}
The contour $\g$. }
\end{center}
\end{figure}

\smallskip
\noindent
{\bf Proof.} 
Using the integral representation of $1/\Gamma(z)$,
\begin{equation}
 \frac{1}{\Gamma(z)} 
 =
 \frac{i}{2\pi} \int_\g (-w)^{-z} e^{-w} dw,
\end{equation}
where $\g$ is the contour in Fig. \ref{figgcontour} 
and (\ref{GinvLap2F}), we see
\begin{align}
 &\quad 
 {\rm rhs ~of}~(\ref{F0dFd}) \notag\\
 &=
 \sum_{n=0}^{\i}
 \frac{\Gamma(\nu)i}{n! 2\pi }\int_{\gamma} dw e^{-w} (-w)^{-\nu} (-\log(-w))^n 
 \frac{1}{2\pi i}\int_{\d+i\R} d\xi \frac{\xi^n e^{\xi y}}{\Gamma(\xi+1)}
 \int_0^{\i} u^{\xi-1} G(u) du \notag\\
 &=
 \frac{\Gamma(\nu)}{2\pi i}\int_{\d+i\R} d\xi \frac{e^{\xi y}}{\Gamma(\xi+1)}
 \int_0^{\i} u^{\xi-1} G(u) du 
 \frac{i}{2\pi }\int_{\gamma} dw e^{-w} (-w)^{-\nu-\xi} \notag\\
 &=
  \frac{1}{2\pi i}\int_{\d+i\R} d\xi \frac{e^{\xi y}\Gamma(\nu)}{\Gamma(\xi+1)\Gamma(\xi+\nu)}
  \int_0^{\i} u^{\xi-1} G(u) du,
\end{align}
which is rhs of (\ref{lp13}). 
\qed

There is also a relation at the level of the Laplace transform. 
From (\ref{G0gG}) with (\ref{gGamma}), we have
\begin{align}
G_0(e^v) = \frac{\Gamma(\a-a)}{\Gamma(\a-a+d/dv)}G(e^v).
\label{lp31}
\end{align}
Expanding formally rhs of the equation above around $d/dv$,
we obtain the relation.
\begin{equation}
 G_0(e^v)
 =
 \Gamma(\a-a) \sum_{n=0}^\i \frac{1}{n!}  
\left.
 \frac{d^n}{d\nu^n} \left( \frac{1}{\Gamma(\nu)} \right)\right|_{\nu\rightarrow \a-a}
 \frac{d^n}{dv^n}G(e^v),
\label{G0Gs}
\end{equation}
where $G_0(u)$ and $G(u)$ are given by (\ref{GG0}). 
The representation  (\ref{lp31}) or (\ref{G0Gs}) can be used to establish the $F_0$ asysmptotics 
for the stationary OY model just as in the case of the stationary $q$-TASEP. 
(See section 5 in~\cite{IS2017p})

\smallskip
Hereafter we explain how we can  get an integral representation for the free energy distribution of the stationary OY model $\log Z_N(\t,\a)$ where
$Z_N(\t,\a)$ is introduced above~\eqref{l122}. 
Recall that as discussed in~\eqref{l133}, 
it has the same distribution as $\lim_{a\rightarrow\a} Z^{(0)}(\t,\a,a)$.
From~\eqref{lp13},
we have
\begin{align}
\P(\log Z_N(\t,\a)\le y)
=
\lim_{a\rightarrow \a} F_0(u)
=
\int_{i\R}\frac{d\xi}{2\pi i}
\frac
{e^{y\xi}}
{\Gamma(\xi)\Gamma(1+\xi)}
\int_0^{\infty}
du
u^{\xi-1}
\hat{G}(u),
\label{l91}
\end{align}
where 
\begin{align}
\hat{G}(u)
=
\lim_{a\rightarrow \a}
\Gamma(\a-a)
G(u).
\label{l92}
\end{align}
On the other hand, if one takes (\ref{F0dF}), we find 
\begin{equation}
\P(\log Z_N(\t,\a)\le y) 
=
\sum_{n=0}^\i \frac{1}{n!} 
\frac{d^{n}}{d\nu^{n}} \left(\frac{1}{\Gamma(\nu)}\right)\Bigr|_{\nu=1} \hat{F}^{(n+1)}(y),
\label{F0dtF}
\end{equation}
where $\hat{F}^{(n)}(y)$ is the $n$th derivative
of $\hat{F}(y)$ and 
\begin{align}
\hat{F}(y) = \lim_{a\to\a} \Gamma (\a-a)F(y).
\label{l93d}
\end{align}
Note that whilst $F(y)$ is a distribution function,
$\hat{F}(y)$ is not expected to be so. One can use any expression of $\hat{F}(y)$,
for instance ~\eqref{invLapF1} and~\eqref{GinvLap2F} with $F(y)$ (resp $G(u)$)
replaced by $\hat{F}(y)$ (resp. $\hat{G}(u)$). We can use also (\ref{inv3}), 
through it is more suitable to write down a formula for the first derivative.
One has
\begin{equation}
 \hat{F}^{(1)}(y) = \int_\R dw e^{y-e^{y-w}}\frac{1}{\pi} \lim_{\e\downarrow 0} {\rm Im}\hat{G}(-e^{-w}+i\e). 
 \label{FGu}
\end{equation} 

Note that although the functions $\hat{G}(u)$ and $\hat{F}(y)$ are defined 
through $a\rightarrow\a$ limit of $G(u)$ and $F(y)$, 
they have a connection directly 
with the partition function of the stationary OY model $Z_N(\t,\a)$
introduced in section~\ref{sOY}:
since $Z(\t,\a,a)=Z^{(0)}(\t,\a,a)e^{\chi}$
where $\chi\sim-\log \Gamma(\nu)$ with $\nu=\a-a$ (see~\eqref{l80}), we find
$G(u)$ and $G_0(u)$ are related as
\begin{align}
G(u)=\int_{\R}dw \frac{(ue^w)^{\nu}e^{-ue^w}}{\Gamma(\nu)}G_0(e^{-w}).
\label{lc620}
\end{align}
Thus noting~\eqref{l133} and~\eqref{FF0}, we have
\begin{align}
\hat{G}(u):=\lim_{\nu\rightarrow 0}\Gamma(\nu)G(u)
=
\int_{\R}dw e^{-ue^w}
\left\langle
e^{-e^{-w}Z_{N}(\t,\a)}
\right\rangle.
\label{lc624}
\end{align}
We can also have the relation about $\hat{F}(u)$ in a similar way:
\begin{align}
\hat{F}(y)=
\int_{\R}dw e^{-e^{w-y}}\P (\log Z_N(\t,\a)\le w).
\label{lc621}
\end{align}

In both expressions~\eqref{l91} and~\eqref{F0dtF} with~\eqref{FGu},
the remaining problem is to estimate $\hat{G}(u)$.
Note that in our approach, we first take the $q\rightarrow 1$ scaling limit 
in section.~\ref{OYbs} and then take the stationary limit $a\rightarrow \a$. 
As another approach, it would be possible to exchange these two limits,
i.e~\eqref{l92} can also be obtained by taking $q\rightarrow 1$ scaling limit for
the Proposition 5.6 in~\cite{IS2017p}.

Under the specialization~\eqref{l121}, the kernel $K(x_1,x_2)$~\eqref{l7}
can be written as
\begin{align}
K(x_1,x_2)=\sum_{l=0}^{N-2}\phi_l(x_1)\psi_l(x_2)
+(\a-a)B_1(x_1) B_2(x_2),
\label{l95}
\end{align}
where
\begin{align}
&\phi_l(x)=
\frac{1}{2\pi i}\oint dv
\frac
{e^{v x-v^2\t/2}}
{v^{l+1}}
\Gamma(1+v)^{N-1}
\frac
{\Gamma(1+v-a)}
{\Gamma(1+\a-v)},
\label{l47}
\\
&\psi_l(x)
=
\frac{1}{2\pi}
\int_{-\infty}^{\infty}dw
e^{-i w x-w^2\t/2}
(iw)^l
\frac
{1}
{\Gamma(1+iw)^{N-1}}
\frac
{\Gamma(1+\a-i w)}
{\Gamma(1+i w-a)},
\label{l48}
\\
&
B_1(x)
=
\frac{1}{2\pi i}\oint dv
\frac
{e^{v x-v^2\t/2}}
{v^{N-1}(v-a)}
\frac
{\Gamma(1+v-a)\Gamma(1+v)^{N-1}}
{\Gamma(1+\a-v)},
\label{l93}
\\
&
B_2(x)
=
\frac{1}{2\pi}
\int_{-\infty}^{\infty}dw
\frac
{e^{-i w x-w^2\t/2}}
{iw-\a}
\frac
{(iw)^{N-1}}
{\Gamma(1+i w)^{N-1}}
\frac
{\Gamma(1+\a-i w)}
{\Gamma(1+i w -a)},
\label{l94}
\end{align}
for $l=0,1,\cdots,N-2$.
Note that although under~\eqref{l121}, $\phi_l(x)$~\eqref{l8} vanishes and
$\psi_l(x)$~\eqref{l9} diverges due to the factors with $\a_j,~j=1,\cdots,N-1$,
their product $\phi_l(x)\psi_l(x)$ converges to 
that of~\eqref{l47} and~\eqref{l48} for $l\le N-2$ and the second factor 
in~\eqref{l95}  for $l=N-1$. The fact that each $\eqref{l8}$ and~\eqref{l9} 
do not go to~\eqref{l47} and~\eqref{l48} respectively reflects from
the scaling~\eqref{l4}. Since in the limit~\eqref{l121}, the number of
finite $\a_j$s is 1, we should replace $N$ in the scaling of $\l^{(k)}_j$ 
in~\eqref{l4} by $M=1$ as stated in the argument above~\eqref{lpp21}.

We further rewrite the relation~\eqref{l5} under the specialization~\eqref{l121},
(cf. (5.27) in~\cite{IS2017p} for the $q$-TASEP) as
\begin{align}
G(u)
&=
\det(1-A_{a,\a}-(\a-a)f_u B_1\otimes B_2),
\notag\\
&=
(\a-a)\det (1-A_{a,\a})
\left(
L_{\a,a}-\int_{\R}dx (A_{\a,a}\rho_{A_{\a,a}}f_uB_1)(x)B_2(x)
\right),
\label{l17}
\end{align}
where $A_{a,\a}(x_1,x_2)=f_u(x_1)\sum_{k=0}^{N-2}\phi_k(x_1)\psi_k(x_2)$, $f_u$ is defined by~\eqref{l6}, $\rho_{A_{\a,a}}=(1-A_{a,\a})^{-1}$ and
\begin{align}
L_{\a,a}
=
\frac
{1}{a-\a}
-
\int_{\R} dx f_u(x) B_1(x) B_2(x).
\label{l96}
\end{align}
Furthermore we decompose $B_1(x)$ and $B_2(x)$ as
(corresponding to~(5.30) in \cite{IS2017p} for $q$-TASEP),
\begin{align}
B_1(x)=B_1^{(1)}(x)+B_1^{(2)}(x),~B_2(x)=B_2^{(1)}(x)+B_2^{(2)}(x),
\label{l23}
\end{align} 
where $B_1^{(1)}(x)$ (resp. $B_2^{(1)}(x)$) is the residue at $x=a$ (resp. $x=-i\a$),
while $B_i^{(2)}(x)$, $i=1,2$ come from remaining contributions,
\begin{align}
B_1^{(1)}(x)
&=
\frac
{e^{a x-a^2\t/2}}
{a^{N-1}}
\frac
{\Gamma(1+a)^{N-1}}
{\Gamma(1+\a-a)},
\label{l24}
\\
B_2^{(1)}(x)
&=
\frac
{e^{-\a x+\a^2\t/2}}
{\Gamma(1+\a)^{N-1}}
\frac
{\a^{N-1}}
{\Gamma(1+\a-a)},
\label{l25}
\\
B_1^{(2)}(x)
&=
\frac{1}{2\pi i}\oint_{|v|<a} dv
\frac
{e^{v x-v^2\t/2}}
{v^{N-1}(v-a)}
\frac
{\Gamma(1+v-a)\Gamma(1+v)^{N-1}}
{\Gamma(1+\a-v)},
\label{l26}
\\
B_2^{(2)}(x)
&=
\frac{1}{2\pi}
\int_{\R-ic}dw
\frac
{e^{-i w x-w^2\t/2}}
{\a-i w}
\frac
{(iw)^{N-1}}
{\Gamma(1+i w)^{N-1}}
\frac
{\Gamma(1+\a-i w)}
{\Gamma(1+i w -a)},
\label{l27}
\end{align}
where $c$ in~\eqref{l27} satisfies $\a<c<\a+1$.
Using these, we write~\eqref{l96} as
\begin{align}
L_{\a,a}
=
\frac
{1}{\a-a}
-
\int_{\R} dx f_u(x) B_1^{(1)}(x) B_2^{(1)}(x)
-
\sum_{\substack{i,j=1\\ (i,j)\neq (1,1)}}^2
\int_{\R}dx f_u(x) B_1^{(i)}(x) B_2^{(j)}(x).
\label{l29}
\end{align}
Here we take the stationary limit $a\rightarrow \a$ in~\eqref{l91} and~\eqref{l92}.
As with Lemma~5.5 in~\cite{IS2017p} for the stationary $q$-TASEP, 
the following lemma is important.
\begin{lemma}
\label{ll4}
Let $L_{\a}=\lim_{a\rightarrow\a}L_{\a,a}$. We have
\begin{align}
L_{\a}=
(N-1)
\left(
\frac{\Gamma'(1+\a)}{\Gamma(1+\a)}
-\frac{1}{\a}
\right)
-2\gamma_E-\a \t-\log u
-
\sum_{\substack{i,j=1\\ (i,j)\neq (1,1)}}^2
 \int_{\R} dx f_u(x) B_{1}^{(i)}(x;\a) B_{2}^{(j)}(x;\a),
\label{ll41}
\end{align}
where $\gamma_E$ is the Euler-Mascheroni constant and
$B_{j}^{(i)}(x;\a),~i,j=1,2$ are $a\rightarrow \a$ limit of~\eqref{l24}-\eqref{l27}.
\end{lemma}

\smallskip
\noindent
{\bf Proof.}
It is easy to see that the last term in~\eqref{l29} goes to the one in~\eqref{ll41}.
The remaining part is to establish
\begin{align}
\lim_{a\rightarrow \a}\frac{1}{\a-a}
-
\int_{\R} dx f_u(x) B_1^{(1)}(x) B_2^{(1)}(x)
=
(N-1)
\left(
\frac{\Gamma'(1+\a)}{\Gamma(1+\a)}
-\frac{1}{\a}
\right)
-2\gamma_E-\a\t-\log u.
\label{l97}
\end{align}
For this purpose, we calculate
\begin{align}
\int_{\R} f_u(x)  B_1^{(1)}(x) B_2^{(1)}(x)
&=
\left(\frac{\a}{a}
\frac
{\Gamma(1+a)}{\Gamma(1+\a)}
\right)^{N-1}
\frac
{e^{(\a^2-a^2)\t/2}}{\Gamma(1+\a-a)^2}
\int_{\R} dx
\frac
{e^{(a-\a)x}}
{1+e^{-x}/u}
\notag
\\
&=
\left(\frac{\a}{a}
\frac
{\Gamma(1+a)}{\Gamma(1+\a)}
\right)^{N-1}
\frac
{e^{(\a^2-a^2)\t/2}}{\Gamma(1+\a-a)^2}
\frac{\pi u^{\a-a}}
{\sin \pi(\a-a)},
\label{ll44}
\end{align}
where in the second equality we use the formula
\begin{align}
\int_{\R}dx\frac{e^{c x}}{1+e^{x}}
=
\frac
{\pi}{\sin \pi c}
\label{ll43}
\end{align}
for $0<\text{Re}~c<1$. 
Noting
\begin{align}
&
\left(\frac{\a}{a}
\frac
{\Gamma(1+a)}{\Gamma(1+\a)}
\right)^{N-1}
=
1
-(N-1)
\left(
\frac{1}{\a}-\frac{\Gamma'(1+\a)}{\Gamma(1+\a)}
\right)(a-\a)+O((a-\a)^2),
\label{ll45}
\\
&
\frac
{1}{\Gamma(1+\a-a)^2}=1-2\gamma_E (a-\a)+O((a-\a)^2),
\label{ll46}
\\
&
e^{(\a^2-a^2)\t/2}u^{\a-a}=1-(\a \t+\log u)(a-\a)+O((a-\a)^2),
\label{ll47}
\\
&
\frac{\pi}{\sin (\pi(\a-a))}
=
\frac
{-1}{a-\a}+O(a-\a),
\label{ll48}
\end{align}
where in~\eqref{ll46}, we used the fact $\Gamma'(1)=-\gamma_E$, we 
arrive at~\eqref{l97}.
\qed

\smallskip
Combining~\eqref{l17} with Lemma~\ref{ll4}, we find a representation for $\hat{G}(u)$~\eqref{l92}.
Thus we obtain two representations of the free energy distribution for the stationary OY model
substituting the representation of $\hat{G}(u)$
into ~\eqref{l91} and~\eqref{F0dtF}:
\begin{theorem}
\label{lt5}
The free energy distribution of the stationary OY model
with parameter $\a$
is given by
\begin{align}\P(\log Z_N(\t,\a)\le y)
&=
\int_{i\R}\frac{d\xi}{2\pi i}
\frac
{e^{y\xi}}
{\Gamma(\xi)\Gamma(1+\xi)}
\int_0^{\infty}
du
u^{\xi-1}
\hat{G}(u)
\notag
\\
&=
\sum_{n=0}^\i \frac{1}{n!} 
\frac{d^{n}}{d\nu^{n}} \left(\frac{1}{\Gamma(\nu)}\right)\Bigr|_{\nu=1} \hat{F}^{(n+1)}(y).
\label{lt59}
\end{align}
Here $\hat{G}(u)$ is given by 
\begin{align}
\hat{G}(u)
&
=
\det (1-A_{\a})
\left(
L_{\a}-\int_{\R}dx (A_{\a}\rho_{A_\a}f_u B_{1})(x;\a)B_{2}(x;\a)
\right),
\label{lt52}
\end{align}
where $A_\a=\lim_{a\rightarrow \a}A_{\a,a}$ with $A_{\a,a}$ defined below~\eqref{l17},
$\rho_{A_{\a}}=(1-A_{\a})^{-1}$,
$B_j(x;\a)=\lim_{a\rightarrow\a}B_j(x)$ for $j=1,2$ and $L_{\a}$ is given 
by~\eqref{ll41}.
In the second expression we can choose several representation, 
for instance ~\eqref{invLapF1} and~\eqref{GinvLap2F} with $F(y)$ (resp. $G(u)$)
replaced by $\hat{F}(y)$ (resp. $\hat{G}(u)$), or~\eqref{FGu} for $\hat{F}^{(1)}(y)$.  
\end{theorem}
\smallskip 
\noindent
{\bf Proof.}
We immediately obtain ~\eqref{lt52}
substituting~\eqref{l17} into rhs of~\eqref{l92} and using Lemma~\ref{ll4}.
\qed

\smallskip
\noindent
Using the relation 
\begin{align}
\det (1-A_{\a})
\left(1-\int_{\R}dx (A_{\a}\rho_{A_\a}f_u B_{1})(x;\a)B_{2}(x;\a)
\right)=\det (1-A_{\a}-(A_{\a}f_u B_{1,\a})\otimes B_{2,\a}),
\label{lt57}
\end{align}
we find~\eqref{lt52} can also be written as
\begin{align}
\hat{G}(u)=\det\left(1-A_{\a}\right)(L_{\a}-1)
+
\det\left(1-A_{\a}-(A_\a f_{u}B_1)\otimes B_2\right).
\label{lt58}
\end{align}

Combining~\eqref{FGu} with~\eqref{lt52}, we obtain the following representation of $\hat{F}(y)$: 
\begin{align}
\hat{F}(y) = \int_\R dw e^{-e^{y-w}}
\left(\hat{G}(-e^{-w})-\hat{G}^{(\delta)}(-e^{-w})\right).
\label{lt56}
\end{align}
Here $\hat{G}^{(\delta)}(-e^{-w})$ is defined by~\eqref{lt52} with
$f_{-e^{-w}}(x)=1/(1-e^{w-x})$ (see~\eqref{l6}) replaced by $1/(1-e^{w-x})-\delta(x-w)$.
For showing~\eqref{lt56}, it is sufficient to show that the part 
$-\frac{1}{\pi} \lim_{\e\downarrow 0}\text{Im}\hat{G}(-e^{-w}+i \e)$
in~\eqref{FGu} is written as
\begin{align}
-\frac{1}{\pi} \lim_{\e\downarrow 0}\text{Im}\hat{G}(-e^{-w}+i \e)
=
\hat{G}(-e^{-w})-\hat{G}^{(\delta)}(-e^{-w}).
\label{lt54}
\end{align}
This is obtained by using the fact $1/(x+i\e)=\mathcal{P}(1/x)\mp i\pi \delta(x)$,
where $\mathcal{P}$ represents the Cauchy principal value and basic properties of 
determinant.

Although  $\hat{G}^{(\delta)}(-e^{-w})$ includes the delta function terms, 
we find that it is finite:
As with the representation~\eqref{lt58} of $\hat{G}(u)$, we can also express 
$\hat{G}^{(\delta)}(-e^{-w})$ as
\begin{align}
\hat{G}^{(\delta)}(-e^{-w})
=
\det\left(1-A_{\a}^{(\delta)}\right)(L^{(\delta)}-1)
+
\det\left(1-A_{\a}^{(\delta)}-(A^{(\delta)}_\a f_{-e^{-w}}^{(\delta)}B_1)\otimes B_2\right),
\label{lt55}
\end{align}
where $L_\a^{(\delta)}=L_\a+B_1(w)B_2(w)$, $A_\a^{(\delta)}(x,y)=
(f_{-e^{-w}}(x)-\delta(x-w))K(x,y)$and $f_{-e^{-w}}^{(\delta)}(x)=f_{-e^{-w}}(x)-\delta(x-w)$.
Note that each integral in the expansion of the above two 
Fredholm determinants is finite even if it has delta function
contributions.

Applying the same arguments and calculations as for the long-time
limit of the $q$-TASEP (see section 5.3 in~\cite{IS2017p}), we finally obtain 
the limiting distribution. 
\begin{corollary}
\label{lc6}
\begin{align}
\lim_{N\rightarrow\i}
\P
\left(
\frac
{\log Z_N(\k N,\a)-\eta N}
{\gamma N^{1/3}}\le s
\right)
=
F_{\rm{BR}}(s;\o),
\label{lc61}
\end{align} 
where $F_{\text{BR}}(s;\o)$ denotes the Baik-Rains distribution and
\begin{align}
\a=\theta-\frac{\o}{\gamma N^{1/3}},
~~
\k=\Phi'(\theta),
~~\eta=\Phi\rq{}(\theta)\theta-\Phi (\theta),
~~\gamma
=\left(
\frac
{-\Phi''(\theta)}
{2}
\right)^{1/3},
\label{lc62}
\end{align}
where $\theta>0$ and $\Phi(z)=\Gamma'(z)/\Gamma(z)$ is the digamma function.
(Note that $\Phi''(z)<0$ for $z>0$.)
\end{corollary}

\smallskip
\noindent
{\bf Remark.}
The Baik-Rains distribution has a few different representations. Here
it is convenient to choose the one in~\cite{IS2013,IS2017p}
\begin{align}
&F_{\text{BR}}(s;\o)=\frac{\partial}{\partial s}\nu_{\o}(s),
\notag\\
&\nu_{\o}(s)
=
F_2(s) 
\left(
s-\o^2-\sum_{\substack{i,j=1\\(i,j)\neq (1,1)}}^2\int_{s}^{\i}d\xi\B_\o^{(i)}(\xi)\B_{-\o}^{(j)}(\xi)
-\int_{s}^{\infty}d\xi
(\rho_{\A}\A\B_{\o})(\xi)\B_{-\o}(\xi)
\right),
\label{fp52}
\end{align}
where $F_2(s)=\det(1-\A)_{L^2(\R)}$ with the Airy kernel 
$\A(x,y)=1_{\ge s}(x)\int_{0}^\infty d\l \Ai (x+\l)\Ai(y+\l)$ 
is the GUE Tracy-Widom distribution~\cite{TW1994} and
\begin{gather}
\rho_{\A}(x,y)=(1-\A)^{-1}(x,y),~~
\B_{\o}^{(1)}(\xi)=e^{\o^3/3-\o\xi},
~~
\B_{\o}^{(2)}(\xi)=-\int_{0}^{\i}dz 
e^{\o z}\Ai(\xi+z),
\notag
\\
\B_{\o}(\xi)=\B_{\o}^{(1)}(\xi)+\B_{\o}^{(2)}(\xi).
\label{fl45}
\end{gather}
\smallskip
\noindent
{\bf Proof.} For showing~\eqref{lc61}, it is convenient to use~\eqref{F0dtF}.
Scaling $\t$ and $y$ as $\t=\k N,~y=\eta N+\gamma N^{1/3}s$  
in both hand sides, we have
\begin{align}
\lim_{N\rightarrow\i}\P(\log Z_N(\t,\a)\le y)
=
\frac{d}{ds}
\lim_{N\rightarrow\i}
\frac{\hat{F}(y)}
{\gamma N^{1/3}}.
\label{lc69}
\end{align}
In order to consider the limit in rhs, we first focus on the relation~\eqref{lc621}.
Associated with the scaling of $y$ stated above~\eqref{lc69}, we scale $w$ in~\eqref{lc621}
also as $w=\eta N+\gamma N^{1/3}x$ then take the limit $N\rightarrow\infty$. 
We have
\begin{align}
\lim_{N\rightarrow\i}\frac{\hat{F}(y)}{\gamma N^{1/3}}
&=
\lim_{N\rightarrow\i}\int_{\R}dx e^{-e^{-\gamma N^{1/3}(s-x)}}
\P\left(\log Z_{N}(\t,\a)\le w\right)
\notag\\
&=
\int_{-\infty}^s dx \lim_{N\rightarrow\i} \P\left(\log Z_{N}(\t,\a)\le w\right),
\label{lc622}
\end{align}
where we used the fact $\lim_{N\rightarrow\i}e^{-e^{-N x}}=1_{\ge 0}(x)$.
Noting the relation 
\begin{align}
\lim_{N\rightarrow\i} \P\left(\log Z_{N}(\t,\a)\le w\right)
=
\lim_{N\rightarrow\i} 
\left\langle
e^{-e^{-w}Z_N(\t,\a)}
\right\rangle
\end{align}
and~\eqref{lc624},
we eventually obtain
\begin{align}
\lim_{N\rightarrow\i}\frac{\hat{F}(y)}{\gamma N^{1/3}}
=
\int_{-\infty}^s dx \lim_{N\rightarrow\i}
\left\langle
e^{-e^{-w}Z_N(\t,\a)}
\right\rangle
=
\lim_{N\rightarrow\i}\frac{\hat{G}(e^{-y})}{\gamma N^{1/3}}.
\label{lc623}
\end{align}

Thus for establishing~\eqref{lc61}, it is sufficient to estimate 
$\lim_{N\rightarrow\i}\hat{G}(e^{-y})/
\gamma N^{1/3}.$ 
As with Lemma 5.10 and 5.11 in~\cite{IS2017p}
for the stationary $q$-TASEP, We show that
under the scaling
\begin{align}
x=\eta N+\gamma N^{1/3}\xi,~l=N-\gamma N^{\frac13}\theta\lambda,
\label{lc630}
\end{align}
we have
\begin{align}
&\lim_{N\rightarrow\i}
C_{N,l,\theta,x} \theta \gamma N^{1/3}\phi_l(x)
=
\lim_{N\rightarrow\i}
C_{N,l\theta,x}^{-1}\gamma N^{1/3} \psi_l(x)=\Ai (\xi+\l),
\label{lc610}
\\
&\lim_{N\rightarrow\i}
C_{N,N-1,\theta,x} B_1^{(i)}(x)=\B_{\o}^{(i)}(\xi),~~
\lim_{N\rightarrow\i}
C_{N,N-1,\theta,x}^{-1} B_2^{(i)}(x)=\B_{-\o}^{(i)}(\xi),~~
\text{for~}i=1,2,
\label{lc611}
\\
&\lim_{N\rightarrow\i}
\frac{1}{\gamma N^{1/3}}\left[(N-1)
\left(
\frac{\Gamma'(1+\a)}{\Gamma(1+\a)}
-\frac{1}{\a}
\right)
-2\gamma_E-\a \t+y\right]=s-\o^2,
\label{lc612}
\end{align}
where $C_{N,l,\theta,x}$ is
\begin{align}
C_{N,l,\theta,x}
=\frac{e^{N(\kappa\theta^2/2-\theta x)}\theta^l}
{\Gamma(1+\theta)^{N-1}}.
\label{lc613}
\end{align}
Here we give a sketch of proofs of these relations. For \eqref{lc610} and~\eqref{lc611},
we only focus on the second one in each relation since the first one can be obtained 
in a parallel way. They can be obtained by the saddle point analyses. 
First we focus
on the result on $\Psi_l(x;\a)$ in~\eqref{lc610}. Setting $z=iw$, one has
\begin{align}
\psi_{l}(x;\a)=\frac{1}{2\pi i}\int_{i\R} dz e^{Nf (z)}e^{-\gamma N^{1/3}\xi z}
z^{l-N}\Gamma(1+z)\frac{\Gamma(1+\a-z)}{\Gamma(1+z-\a)},
\label{lc614}
\end{align}
where $f(z)=\k z^2/2-\eta z-\log\Gamma(z)$. One easily finds $z=\theta$ is a double
saddle point i.e. $f\rq{}(\theta)=f\rq{}\rq{}(\theta)=0$. Thus scaling $z$ around
this double saddle point as $z=z(\sigma)=\theta-i\sigma/\gamma N^{1/3}$,
we get
\begin{align}
Nf(z)=Nf(\theta)+\frac{i}{3}\sigma^3+O\left(N^{-1/3}\right).
\label{lc615}
\end{align}
Combining this with the relations
\begin{align}
\lim_{N\rightarrow\infty}e^{-\gamma N^{1/3}\xi (z-\theta)}=e^{i\xi\sigma},~
\lim_{N\rightarrow\infty}\theta^{\theta\gamma N^{1/3}\l}z^{l-N}=e^{i\lambda\sigma},~
\lim_{N\rightarrow\i}
\frac{\Gamma(1+z)\Gamma(1+\a-z)}
{\Gamma(1+\theta)\Gamma(1+z-\a)}
=1,
\label{lc616}
\end{align}
we obtain the second relation in~\eqref{lc610}.

Next we consider the limit of $B_2^{(1)}(x;\a)$ and $B_2^{(2)}(x;\a)$.
The former one can be written as 
\begin{align}
B_2^{(1)}(x;\a)=e^{N f(\a)}e^{-\a\gamma N^{1/3}\xi}\Gamma(\a).
\label{lc617}
\end{align}
Here the function $f(z)$ is defined below~\eqref{lc614}. As with~\eqref{lc615}
and~\eqref{lc616} noting  
$Nf(\a)=Nf(\theta)-\o^3/3+O(N^{-1/3})$, $e^{-\a\gamma N^{1/3}\xi}
=e^{-\theta\gamma N^{1/3}\xi+\xi\o}$, we arrive at the second relation
in~\eqref{lc611} with $i=1$. Also we rewrite $B_2^{(2)}(x;\a)$ as  
\begin{align}
B_2^{(2)}(x;\a)
=\frac{1}{2\pi i}\int_{i\R-c} dz \frac{e^{Nf (z)}}{\a-z}e^{-\gamma N^{1/3}\xi z}
\Gamma(z)\frac{\Gamma(1+\a-z)}{\Gamma(1+z-\a)}
\label{lc618}
\end{align}
with $\a<c<\a+1$. Changing $z$ to $\sigma$ defined above~\eqref{lc615},
and using~\eqref{lc615} and~\eqref{lc616} with the
relation
\begin{align}
\frac{1}{\gamma N^{1/3}(\a-z)}=\frac{1}{i\sigma-\o}
=
-\int_{0}^{\infty}dz e^{(i\sigma-\o)z},
\label{lc619}
\end{align}
where we used the fact that 
the contour of $\sigma$ become $\R-i c$ with $c<\o$,
we obtain the second relation in~\eqref{lc611} with $i=2$.

For the last one~\eqref{lc612}, noting $\Gamma'(1+\a)/\Gamma(1+\a)
-1/\a=\Phi (\a)$ where
$\Phi(\a):=\Gamma'(\a)/\Gamma(\a)$ is the digamma function,
and expand each term in lhs up to $O(N^{1/3})$, we obtain rhs.

These relations with $\lim_{N\rightarrow\infty}f_{e^{-y}}(x)=1_{\ge s}(\xi)$
under the scaling of $x$ and $y$ stated~\eqref{lc630} and above~\eqref{lc69}
lead to
\begin{align}
\lim_{N\rightarrow\i}
\frac{\tilde{G}(e^{-y})}{\gamma N^{1/3}}=\nu_{\o}(s).
\label{lc614d}
\end{align}
Thus from this equation with~\eqref{lc69} and~\eqref{lc623}, we obtain~\eqref{lc61}.
\qed

\section{The stationary KPZ equation}
In this section we consider the limit to the KPZ equation,
\begin{equation}
\partial_t h = \tfrac12\partial_x^2 h + \tfrac12(\partial_x h)^2 + \eta , 
\label{KPZeq}
\end{equation} 
where $h=h(x,t)$ represents the height at position $x\in\R$ and at time $t\geq 0$ and 
$\eta=\eta(x,t)$ is the space-time Gaussian white noise with mean zero and covariance 
$\langle \eta(x,t)\eta(x',t')\rangle = \delta(x-x')\delta(t-t')$,
especially for the stationary situation. It has been known that in the stationary 
KPZ equation, the height difference $h(x,t)-h(0,t)$ is given by the two-sided
Brownian motion. Thus we prepare the initial condition as
\begin{align}
h(x,0)(=h(x,0)-h(0,0))=\tilde{B}(x),
\label{l110}
\end{align} 
where in lhs, we set $h(0,0)=0$ due to the translational invariance and
$\tilde{B}(x)$ is the two-sided Brownian motion with drift $v$~\eqref{l131}.
The KPZ equation is (formally) transformed to the stochastic heat equation (SHE). 
By applying the Cole-Hopf transformation $\Zm(x,t)=e^{h(x,t)}$, $\Zm(x,t)$ 
solves the stochastic heat equation (SHE)
\begin{align}
&\frac{\partial}{\partial t}\Zm(x,t)=\frac{1}{2}\frac{\partial^2}
{\partial x^2}\Zm(x,t)+\eta(x,t)\Zm(x,t),
\notag
\\
&\Zm(x,0)=e^{\tilde{B}(x)}.
\label{SHE}
\end{align}
A precise meaning of the KPZ equation (\ref{KPZeq}) 
for the case of the initial data with the two-sided Brownian motion
is given for example using the Cole-Hopf transformation \cite{BG1997}. 

Now let us consider the scaling limit of the stationary OY model $Z_N(\t,\a)$ in~\eqref{l122}
to $\Zm(x,t)$ in \eqref{SHE}. 
The scaling limit of the OY model 
or more generalized discrete models to the SHE has been studied 
in~\cite{CT2015p}. In our case, we scale $\t$ and $\a$ as
\begin{align}
\t=\sqrt{tN}+x,~
\a
=
\sqrt
{\frac{N}{t}}
+
\frac12+v,
\label{l31}
\end{align}
and set
\begin{align}
C(N,x,t):=\exp\left(N+\frac{\sqrt{tN}+x}{2}+x\sqrt{\frac{N}{t}}\right)
\left(\frac{t}{N}\right)^{\frac{N-1}{2}}.
\label{l33}
\end{align}
Then $Z_{N}(\t,\a)$
goes to
the solution for the stationary SHE~\eqref{SHE},
\begin{align}
\lim_{N\rightarrow\infty}
\frac
{Z_{N}(\t,\a)}
{C(N,x,t)}
=\Zm (x,t).
\label{l32}
\end{align}
Here we give a derivation
of~\eqref{l32}
based on the discussion in section 5.1 in~\cite{IS2016}.

Let $\tilde{Z}_{j,\beta}(\t),~j\in\{1,2,\cdots,N\}$ be
\begin{align}
\tilde{Z}_{j,\beta}(\t)
=
\frac
{e^{-\t-\beta^2\t/2}Z_{j}(\beta^2\t,\a)}
{\beta^{2(j-1)}}.
\label{l124}
\end{align}
Noting $Z_{j}(\t,\a)$~\eqref{l122} solves~\eqref{l130d} 
where $B_1(\t)$ is the standard Brownian motion with drift $\a$ while
the other ones $B_2(\t),\cdots,B_N(\t)$ are the independent standard
Brownian motions without drift, we see that the deformed one~\eqref{l124} satisfies the stochastic differential equation
\begin{align}
d\tilde{Z}_{j,\beta}(\t)=\left(\tilde{Z}_{j-1,\beta}(\t)-\tilde{Z}_{j,\beta}(\t)\right)d\t+\beta \tilde{Z}_{j,\beta}(\t)
dB_j(\t),
\label{ddf}
\end{align}
where we set $\tilde{Z}_{0,\beta}(t)=0$. Now let us take the 
diffusion scaling for~\eqref{ddf}: we set
\begin{align}
\t=tM,~~
j=tM-x\sqrt{M}
\label{KPZs1}
\end{align}
with $M>0$.
Note that at this stage the scaling is different from~\eqref{l31}.
At the same time we scale $\beta$ as
\begin{align}
\beta=M^{-1/4},
\label{1/41}
\end{align}
then take the large $M$ limit. 
Under this scaling limit, we see that~\eqref{ddf} goes to SHE i.e. the first
equation in~\eqref{SHE}. An explanation of this property was given
in section 5.1 in~\cite{IS2016} .

Next we consider the initial condition. Considering $\t=0$ in~\eqref{KPZs1}, 
we notice that only the negative region $x<0$ appears.
This comes from the replacement of the whole $Z_j(\t,\a)$ with the negative
 index $j\le 0$ by a single $Z_1(\t,\a)$ with the Brownian motion 
with drift $\a$ (see the explanation above~\eqref{l122}). Here in order to take 
the region $x>0$ into account, we consider $Z_{j}(0,\a)$ with $j\in\Z$
rather than $j\in\{1,\cdots,N\}$. According to Theorem 3.3 in~\cite{SeVa2010}, one has
\begin{align}
\tilde{Z}_{j,\beta}(0)
=
\frac{Z_{j}(0,\a)}
{\beta^{2(j-1)}}
\sim
\begin{cases}
e^{\sum_{k=1}^{j-1}(r_k(0)-\log\b^2)},& j\ge 0,
\\
e^{\sum_{k=1}^{-j-1}(r_{-k}(0)-\log\b^2)},& j\le -1,
\end{cases}
\label{l125}
\end{align}
where $r_k(0),~k\in \Z$ are i.i.d. random variables
with 
$r_k(0)\sim-\log\Gamma (\a)$ (see below~\eqref{l130d}).
Here we scale $j$ as~\eqref{KPZs1} with $t=0$
and
\begin{align}
\a=\beta^{-2}+v+\frac{1}{2}.
\label{l126}
\end{align}
From the properties of the distribution $-\log\Gamma(\a)$,
we see that
\begin{align}
&\E(r_k(0))=-\log\Gamma\left(\a\right)'
=-\log\a+\frac{1}{2\a}
+o\left(\frac{1}{\a}\right)
=\log \beta^2-\frac{v}{M^{1/2}}+o\left(\frac{1}{M^{1/2}}\right),
\label{l127}
\\
&\text{Var} (r_k(0))
=
\log\Gamma\left(\a\right)''
=
\frac{1}{\a}+o\left(\frac{1}{\a}\right)
=
\frac{1}{M^{1/2}}+o\left(\frac{1}{M^{1/2}}\right).
\label{l128}
\end{align}
Using them and Donsker's theorem~\cite{KaratzasShreve1991}, we 
find that
\begin{align}
&\lim_{M\rightarrow\i}\sum_{k=1}^{j-1}(r_k(0)-\log\beta^2)
=
B_-(x)+vx,~\text{for}~j\ge0 \text{~and~} x\le 0,
\notag
\\
&\lim_{M\rightarrow\i}\sum_{k=1}^{-j-1}(r_{-k}(0)-\log\beta^2)
=
B_+(x)+vx,~\text{for}~j<0 \text{~and~} x> 0,
\label{l129}
\end{align}
where $B_{\pm}(x)$ are independent standard Brownian motion without drift.

Therefore under the scaling~\eqref{KPZs1},~\eqref{1/41} and~\eqref{l126}, the following limiting property is
established.
\begin{align}
\lim_{M\rightarrow\infty}\tilde{Z}_{j,\beta}(\t)
=\Zm(x,t).
\label{CSHE1}
\end{align}

At last we show that the relation~\eqref{l32} with~\eqref{l31} and~\eqref{l33}
is equivalent to~\eqref{CSHE1} with~\eqref{KPZs1},~\eqref{1/41}, and~\eqref{l126}.
We note that even if we slightly change the scaling~\eqref{1/41} to
$\beta=(M-x\sqrt{M}/t)^{-1/4}$, we have the same result~\eqref{CSHE1}.
Setting $N=tM-x\sqrt{M}$, one sees that~\eqref{KPZs1} and~\eqref{l126} with the modification above is equivalent to
\begin{align}
\t=N+x\sqrt{\frac{N}{t}},~
j=N,~
\beta=\left(\frac{N}{t}\right)^{-1/4},~
\a=\left(\frac{N}{t}\right)^{1/2}+v+\frac{1}{2}.
\label{KPZs2}
\end{align}
Applying the above scaling to rhs of~\eqref{l124}
and noticing
\begin{align}
\beta^2\t=\sqrt{tN}+x,~
\beta^{2(N-1)}e^{\t+\beta^2\t/2}
=C(N,x,t),
\end{align}
where $C(N,x,t)$ is defined in~\eqref{l33}, we find that~\eqref{CSHE1} 
is equivalent to~\eqref{l32}.

The goal of this section is to obtain the height distribution function of the stationary
KPZ equation~\eqref{KPZeq} and~\eqref{l110} by considering the scaling limit of Theorem~\ref{lt5} in the stationary OY model. Hereafter we put a 
tilde ($\tilde{\quad}$) on each $\hat{G},\hat{F},~A_{\a}$ and $B_{i,\a}(x),~i=1,2$ in~\eqref{lt59} and~\eqref{lt52}, for the quantities for the OY model while those without tildes represent
the corresponding quantities for the KPZ equation~\eqref{KPZeq}.

We define $\hat{G}(u)$ and $\hat{F}(s)$ as
\begin{align}
&\hat{G}(u)=\int_{\R} dw e^{-u e^{w}}
\left\langle
\exp\left(-e^{h(2\gamma_t^2y,t)+\frac{\gamma_t^3}{12}+\gamma_t y^2-w}
\right)
\right\rangle,
\label{l176}
\\
&\hat{F}(s)=
\int_{\R}dw e^{-e^{w-s}}
\P
\left(
h(2\gamma_t^2 y,t)+\frac{\g_t^3}{12}+\gamma_t y^2
\le w
\right),
\label{l170}
\end{align}
where we set
\begin{align}
x=2\gamma_t^2 y,~v=\o/\gamma_t,~\gamma_t=(t/2)^{1/3}.
\label{l200}
\end{align}
Note that these can be obtained as the KPZ equation limit of 
$\tilde{\hat{G}}(\tilde{u})$~\eqref{lc624} and $\tilde{\hat{F}}(\tilde{y})$~\eqref{lc621} 
respectively for the stationary OY model, i.e.
in addition to~\eqref{l31} with~\eqref{l200}, we scale $\tilde{u}$ in $\tilde{\hat{{G}}}(\tilde{u})$
and $\tilde{y}$ in $\tilde{\hat{{F}}}(\tilde{y})$ as
\begin{align}
\tilde{u}=
ue^{\frac{\gamma_t^3}{12}+\gamma_t y^2}/
C(N,t,2\gamma_t^2y),~\tilde{y}=s -\frac{\gamma_t^3}{12}-\gamma_t y^2+
\log C(N,t,2\gamma_t^2y)
\label{ll75}
\end{align}
respectively.
Then under the above scaling we have
\begin{align}
\hat{G}(u)=\lim_{N\rightarrow\infty} \tilde{\hat{G}}(\tilde{u}),~~
\hat{F}(s)=\lim_{N\rightarrow\infty} \tilde{\hat{F}}(\tilde{y}).
\label{l171}
\end{align}

Before stating result, we further define some functions:
\begin{align}
&\Ai_{\Gamma} (\xi,\o)
=\frac{1}{2\pi}
\int_{\Gamma_{\o}} dz 
e^{iz \xi+i\frac{z^3}{3}}
\frac
{\Gamma\left(1-(\o-iz)/\g_t\right)}
{\Gamma\left(1+(\o-iz)/\g_t\right)},
\label{l36}
\\
&B_{\o}^{(1)}(\xi)=
e^{-\o^3/3+\o \xi},
~B_{\o}^{(2)}(\xi)=-\int_{0}^{\infty}d\lambda e^{-\o\lambda}\Ai_{\Gamma}
\left(\xi+\lambda, \o\right),
\label{l38}
\\
&B_\o(\xi)=B_\o^{(1)}(\xi)+B_\o^{(2)}(\xi).
\end{align}
In~\eqref{l36}, $\Gamma_\o$
represents the contour from $-\i$ to $\i$ passing
below the pole $i(\gamma_t-\o)$. 
Then we have the following.
\begin{proposition}
\label{lp6}
We have the following representation for
$\hat{G}(u)$~\eqref{l176}.
\begin{align}
\hat{G}(u)
&
=
\gamma_t
\det (1-A_{\o+y,u})
\left(
L_{\o+y,u}-\int_{\R}d\xi (A_{\o+y,u}\rho_{A_{\o+y,u}}f_{u,\gamma_t}B_{\o+y})(\xi)
B_{-(\o+y)}(\xi)
\right),
\label{l34}
\end{align}
where $f_{u,\gamma_t}(\xi)=f_u(\gamma_t\xi)$
and
\begin{align}
&L_{\o,u}
=
-
\frac{2\gamma_E+\log u}{\g_t}- \o^2
-
\sum_{\substack{i,j=1\\ (i,j)\neq (1,1)}}^2
\int_{\R} d\xi f_{u}(\gamma_t \xi) B_{\o}^{(i)}(\xi) B_{-\o}^{(j)}(\xi),
\label{l35}
\\
&A_{\o,u}(\xi_1,\xi_2)
=f_{u}(\gamma_t \xi_1)
\int_{0}^{\infty}d\lambda
\Ai_{\Gamma}
\left(\xi_1+\l,\o\right)
\Ai_{\Gamma}
\left(\xi_2+\l,-\o\right).
\label{l37}
\end{align}
\end{proposition}

\smallskip
For the proof this proposition, the following lemma plays a crucial role.
The proof will be given in Appendix~\ref{proofll7}
\begin{lemma}
\label{ll7}
In addition to~\eqref{l31} with~\eqref{l200}, 
we scale $\tilde{u}$ as the first one in~\eqref{ll75} 
and $\tilde{x}$, the argument of the functions
$\Phi(\tilde{x}),~\Psi(\tilde{x}),~B^{(i)}_j(\tilde{x}),~i,j=1,2$,
 and $l$ as
\begin{align}
\tilde{x}=
\gamma_t \xi -\frac{\gamma_t^3}{12}-\gamma_t y^2+
\log C(N,t,2\gamma_t^2 y),
~l=N-\sqrt{\frac{N}{2\gamma_t}}\l.
\label{ll76}
\end{align}
Then under~\eqref{l31} and~\eqref{ll76}, we have
\begin{align}
&\lim_{N\rightarrow\infty}
\frac{\sqrt{N}e^{f_N(\tilde{z};\t,\tilde{x},l)}\Gamma(1+\sqrt{N}\tilde{z})}{(2\gamma_t)^{1/2}}
\Phi_l(\tilde{x})
=
\Ai_{\Gamma}
\left(\xi+\l,\o+y\right),
\notag\\
&
\lim_{N\rightarrow\infty}
\frac{1}{e^{f_N(\tilde{z};\t,\tilde{x},l)}\Gamma(1+\sqrt{N}\tilde{z})}
\Psi_l(\tilde{x})
=
\Ai_{\Gamma}
\left(\xi+\l,-\o-y\right),
\label{ll71}
\\
&
\lim_{N\rightarrow\i}
e^{f_N\left(z_c+\frac{1}{2\sqrt{N}};\t,\tilde{x},N-1\right)}\Gamma(1+\sqrt{N/t})
\tilde{B}_1^{(1)}(\tilde{x})
=B^{(1)}_{\o+y}(\xi),
\notag
\\
&\lim_{N\rightarrow\i}
\frac
{1}
{e^{f_N\left(z_c+\frac{1}{2\sqrt{N}};\t,\tilde{x},N-1\right)}\Gamma(1+\sqrt{N/t})}
\tilde{B}_2^{(1)}(\tilde{x})
=B^{(1)}_{-\o-y}(\xi),
\label{ll72}
\\
&
\lim_{N\rightarrow\i}
e^{f_N\left(z_c+\frac{1}{2\sqrt{N}};\t,\tilde{x},N-1\right)}\Gamma(1+\sqrt{N/t})
\tilde{B}_1^{(2)}(\tilde{x})
=B^{(2)}_{\o+y}(\xi),
\notag
\\
&
\lim_{N\rightarrow\i}
\frac
{1}
{e^{f_N\left(z_c+\frac{1}{2\sqrt{N}};\t,\tilde{x},N-1\right)}\Gamma(1+\sqrt{N/t})}
\tilde{B}_2^{(2)}(\tilde{x})
=B^{(2)}_{-\o-y}(\xi),
\label{ll73}
\\
&
\lim_{N\rightarrow\i}(N-1)
\left(
\frac{\Gamma'(1+\a)}{\Gamma(1+\a)}
-\frac{1}{\a}
\right)
-2\gamma_E-\a\t-\log\tilde{u}
=-2\gamma_E-\log u-\gamma_t (\o+y)^2,
\label{ll74}
\end{align}
where $f_N(z;\t,\tilde{x},l)$ is defined by~\eqref{fN} and $\tilde{z}$
is given as~\eqref{ll76d} with $\sigma=0$.
\end{lemma}

\smallskip
\noindent
{\bf Proof. of Proposition~\ref{lp6}.}
Combining the first relation in~\eqref{l171} with Lemma~\ref{ll7},
we readily get~\eqref{l34}
\qed

\smallskip

Thus we arrive at an expression of 
the hight distribution for the stationary KPZ equation:
\begin{theorem}
\label{lt6}
Set $\gamma_t=(t/2)^{1/3}$. For $y\in\R, s\in\R$, we have
\begin{align}
\P
\left(
\frac
{h(2\gamma_t^2 y,t)+\frac{\g_t^3}{12}+\gamma_t y^2}
{\g_t}
\le s
\right)
&=
\frac
{\gamma_t}{2\pi i}
\int_{i\R} d\xi
\frac
{e^{\gamma_t s\xi}}{\Gamma(\xi)\Gamma(1+\xi)}
\int_{\R} dw e^{-\gamma_t w\xi}\hat{G}
\left(e^{-\gamma_t w}\right)
\notag\\
&=
\sum_{n=0}^{\i} \frac{1}{\gamma_t^{n+1} n!} 
\frac{d^{n}}{d\nu^{n}}\left( \frac{1}{\Gamma(\nu)}\right) \Big|_{\nu=1} 
\frac
{d^{n+1}}
{d s^{n+1}}
\hat{F}(\gamma_t s).
\label{lt61}
\end{align}
Here $\hat{G}(u)$ is given by~\eqref{l34}.
For $\hat{F}(s)$ in the second expression, which is defined in~\eqref{l170},
one can use the inversion formulas in Appendix~\ref{sinvLap},  
for example,~\eqref{invLapF1} or~\eqref{GinvLap2F} with $F(y)$
and $G(u)$ replaced by $\hat{F}(y)$ and $\hat{G}(u)$. 
One can also use~\eqref{FGu} in which $\hat{F}(y)$ and $\hat{G}(u)$ 
are replaced by those for the KPZ equation~\eqref{l170},\eqref{l176}.
\end{theorem}
\smallskip
\noindent
{\bf Proof of Theorem~\ref{lt6}.}
We take the KPZ equation limit (the limit $N\rightarrow\infty$
under the scaling \eqref{l31} with~\eqref{l200} and~\eqref{ll75} 
with $u$ and $s$ replaced by
$e^{-\gamma_t w}$ and $\gamma_t s$) for~\eqref{lt59}.
(Recall that we put tilde on $u$ and $y$ in~\eqref{lt59}.)
We have
\begin{align}
\P
\left(
\frac
{h(2\gamma_t^2 y,t)+\frac{\g_t^3}{12}+\gamma_t y^2}
{\g_t}
\le s
\right)
&=
\frac
{\gamma_t}{2\pi i}
\int_{i\R} d\xi
\frac
{e^{\gamma_t s\xi}}{\Gamma(\xi)\Gamma(1+\xi)}
\int_{\R} dw e^{-\gamma_t w\xi}
\lim_{N\rightarrow\infty} \tilde{\hat{G}}(\tilde{u})
\notag\\
&=
\sum_{n=0}^{\i} \frac{1}{\gamma_t^{n+1} n!} 
\frac{d^{n}}{d\nu^{n}}\left( \frac{1}{\Gamma(\nu)}\right) \Big|_{\nu=1} 
\frac
{d^{n+1}}
{d s^{n+1}}
\lim_{N\rightarrow\infty} \tilde{\hat{F}}(\tilde{y}).
\end{align}
Combining this with~\eqref{l171}, we immediately obtain~\eqref{lt61}.
\qed

\smallskip
The height distribution for the stationary KPZ equation has already been studied in 
~\cite{IS2012,IS2013} using the replica 
method and by~\cite{BCFV2015} using the Macdonald process technique. 
The first expression in (\ref{lt61}) is close to the one in~\cite{BCFV2015}.
The only difference is that $\gamma_t^{-1}\hat{G}(e^{-\gamma_t w})$~\eqref{l34}
is replaced by $\Xi(S,b,\sigma)$ of (2.11) in  \cite{BCFV2015}
with $S=e^{-\gamma_t w}$, $b=(\o+y)/\gamma_t$, and $\sigma=1/\gamma_t$.
This comes from the choice of the Fredholm determinant representations.
In this paper, we used the formula~\eqref{l5} where under 
the specialization~\eqref{l121}, the kernel $f_u(x_1)K(x_1,x_2)$ 
is expressed as a product of~\eqref{l6} and~\eqref{l95}.
On the other hand in~\cite{BCFV2015}, the authors take the KPZ equation
limit before taking the stationary limit $\tilde{a}\rightarrow\tilde{\a}$, i.e. they
consider the KPZ equation
with the initial condition $h(x,0)=1_{x>0}(B_++a x)+1_{x\le 0}(B_-+\a x)$
in place of~\eqref{l110}, where $B_{\pm}(x)$ are the independent standard 
Brownian motions and obtained a  
Fredholm determinant formula with the kernel
( see (2.5) and (2.6) in~\cite{BCFV2015},) 
\begin{align}
K_{\a,a}(x,y)=
\frac{\gamma_t}{(2\pi i)^2}
\int dw\int dz
\frac{\pi e^{s\gamma_t (z-w)}}{\sin\pi(z-w)}
\frac{e^{(\gamma_t z)^3/3-\gamma_t zy}}{e^{(\gamma_t w)^3/3-\gamma_t wx}}
\frac{\Gamma(\a-z)}{\Gamma(z-a)}
\frac{\Gamma(w-a)}{\Gamma(\a-w)},
\label{l138}
\end{align}
where $x,y\in\R_+$ and $z$ and $w$ satisfy $0<\text{Re}(z-w)<1/2$
(for more precise information  of the contours see Theorem 2.7 in~\cite{BCFV2015}).
One feature of our kernel $f_u(x_1)K(x_1,x_2)$ 
is that the function $f_u(x_1)$ is completely separated
from $K(x_1,x_2)$. This enables us to have a simple rank 1 perturbation
of $K(x_1,x_2)$~\eqref{l95}. On the other hand in~\eqref{l138}, 
the information of $f_u(x)$ is included in 
the factor $\pi e^{\gamma_t s(z-w)}/\sin\pi(z-w)$ as one sees from (cf ~\eqref{ll43})
\begin{align}
\frac{\pi e^{\gamma_t s(z-w)}}{\sin\pi(z-w)}
=
\int_\R dx f_{e^{-\gamma_t s}}(x)e^{-(z-w)x}.
\label{l139}
\end{align}
In the form of (\ref{l138}), it does not seem clear how one can find a simple rank 1 perturbation 
of the kernel, but one can still calculate a rank three perturbation of the kernel
(see section 7 in~\cite{BCFV2015}), from which Proposition 2.14 in~\cite{BCFV2015}
follows.

On the other hand the other expression in~\cite{IS2013} is obtained from
the second expression in~\eqref{lt61} where
\begin{align}
\hat{F}(z) = \int_\R dw e^{-e^{-\gamma_t w}/z} 
\left(
\hat{G}(-e^{-\gamma_t w})-\hat{G}^{(\delta)}(-e^{-\gamma_t w})
\right),
\label{lt62}
\end{align}
and $\hat{G}^{(\delta)}(-e^{-\gamma_t w})$ is defined 
in the same way as $\hat{G}(-e^{-\gamma_t w})$~\eqref{l34}
with $f_{-e^{-\gamma_t w}}(\gamma_t\xi)=1/(1-e^{\gamma_t(w-\xi)})$
replaced by $1/(1-e^{\gamma_t(w-\xi)})-\delta(\xi-w)$.
As discussed in~\eqref{lt58} below, we find that  $\hat{G}^{(\delta)}(-e^{-\gamma_t w})$ 
is finite even if it includes the delta function term.

Finally, in the large $t$ limit, our formula~\eqref{lt61} goes to the distribution 
$F_{\rm BR}(s;\o)$ which was introduced in \cite{BR2000}. 
(See remark below Corollary~\ref{lc6}.)
This can be easily seen by taking the $t\rightarrow\infty$ limit of
the second relation of~\eqref{lt61}, which has been done in
section 6 in~\cite{IS2013}.

\appendix
\section{Two-sided $q$-Whittaker processes}
In this appendix, we summarize basic definitions and properties of the 
two-sided $q$-Whittaker function. For more details, see \cite{IS2017p}. 
\subsection{Definitions}
\begin{figure}[t]
\begin{center}
\includegraphics[scale=1.0]{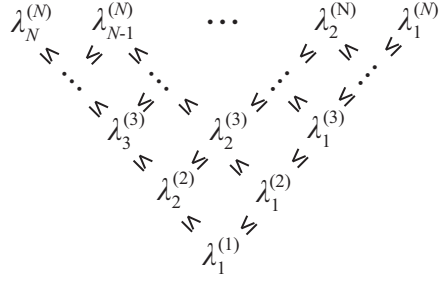}
\caption{\label{figGT}
The Gelfand-Tsetlin cone as a triangular array. 
}
\end{center}
\end{figure}

The set of $n$-tuples of non-increasing integers, each of which 
can take both positive and negative value is denoted by 
\begin{align}
\S_n := 
\{\l=
(\l_1,\cdots,\l_n)\in\Z^n
|\l_1\ge \l_2\ge\cdots\ge\l_n
\}. 
\label{signature}
\end{align}
An element $\l\in\S_n$ is called a signature. 
The set of $N(N+1)/2$-tuples of integers with interlacing conditions, 
\begin{align}
\mathbb{G}_N
:=
&\{ (\l^{(1)},\l^{(2)},\cdots,\l^{(N)}), \l^{(n)}\in\S_n,1\leq n\leq N
| \notag\\
&\quad \l^{(m+1)}_{\ell+1}\le \l^{(m)}_{\ell}\le \l^{(m+1)}_{\ell}, 
1\le \ell \le m \le N-1
\}
\label{GT}
\end{align}
is called the Gelfand-Tsetlin cone for signatures. 
See Fig.  \ref{figGT}.
An element of $\underline{\l}_N\in\GG_N$ can also be regarded as a
point in $\Z^{N(N+1)/2}$
with the above interlacing conditions. 

Next we explain the (skew) $q$-Whittaker function labeled by signatures. 
\begin{definition}{\label{d1}}
Let 
$\l\in\S_n,\mu\in\S_{n-1}$ be two signatures of length $n$ and $n-1$ respectively 
and $a$ be an indeterminate. 
The skew $q$-Whittaker function (with one variable) is defined as 
\begin{align}
 P_{\l/\mu}\left(a \right)
 =\prod_{i=1}^n a^{\l_i}\cdot
	\prod_{i=1}^{n-1}\frac{ a^{-\mu_i} (q;q)_{\l_i-\l_{i+1}}}
        {(q;q)_{\l_i-\mu_i}(q;q)_{\mu_i-\l_{i+1}}} .
\label{skewWh}
\end{align} 
Using this, for a signature $\l\in\S_N$ and $N$ indeterminates 
$a=(a_1,\cdots,a_N)$,
we define the $q$-Whittaker function with $N$ variables  as  
\begin{equation}
P_\l\left(a\right)
=
\sum_{\substack{\l_i^{(k)}, 1\leq i\leq k\leq N-1\\
            \l_{i+1}^{(k+1)} \leq \l_i^{(k)} \leq \l_i^{(k+1)}}}
	\prod_{j=1}^N P_{\l^{(j)}/\l^{(j-1)}}\left(a_j\right).
	\label{Wh}
\end{equation}
Here the sum is over the Gelfand-Tsetlin cone $\GG_N$ with the condition $\l^{(N)}=\l$.
\end{definition}
\noindent
We also define another function labeled by a signature. 
\begin{definition}{\label{d2}}
For a signature $\l$ of length $N$~\eqref{signature}, 
$t>0$
and 
$\a=(\a_1,\cdots,\a_N)\in [0,1)^N$, we define
\begin{align}
Q_\l\left(\alpha,t\right)
 =\prod_{i=1}^{N-1}(q^{\l_i-\l_{i+1}+1};q)_{\infty}
\int_{\T^N}\prod_{i=1}^N\frac{dz_i}{z_i}\cdot P_\l \left(1/z\right)
 \Pi\left(z;\a,t\right)m_N^q\left(z\right), 
\label{defQ}
\end{align}
where $z=(z_1,\cdots,z_N)$ and $1/z=(1/z_1,\cdots,1/z_N)$
are shorthand notations, 
\begin{equation}
m_N^q(z)=\frac{1}{(2\pi i)^NN!}\prod_{1\le i<j\le N}(z_i/z_j;q)_{\infty}(z_j/z_i;q)_{\infty}
\label{Sk}
\end{equation} 
is the $q$-Sklyanin measure, 
\begin{align}
\Pi\left(z;\a,t\right)=\prod_{i,j=1}^N\frac{1}{(\a_i/z_j;q)_{\infty}}
\cdot
\prod_{j=1}^Ne^{z_j t}.
\label{d14}
\end{align}
\end{definition}

Using Definitions (\ref{skewWh}) and (\ref{defQ}), we introduce a measure on $\GG_N$. 
\begin{definition}
 For $\underline{\l}_N\in\GG_N$, we define
\begin{align}
P_t (\underline{\l}_N)
 := 
 \frac{\prod_{j=1}^N P_{\l^{(j)}/\l^{(j-1)}} \left( a_j\right) \cdot Q_{\l^{(N)}} \left(\a,t\right)}{\Pi(a;\a,t)} .
\label{2qWh-pr}
\end{align}
\end{definition}
\noindent
We call this the two-sided $q$-Whittaker process.  
Furthermore using~\eqref{Wh}, the marginal distribution of $P_t (\underline{\l}_N)$
on $\l^{(N)}\in\S_N$ can be written as
\begin{proposition}
For $\l\in\S_N$, we have
\begin{align}
\P(\l^{(N)}=\l)
=
\frac{P_{\l}(a)Q_{\l}(\a,t)}
{\Pi(a;\a,t)}.
\label{2qWh-me}
\end{align}
\end{proposition}

\smallskip
\noindent
We call this the two-sided $q$-Whittaker measure.  

One of the main results in \cite{IS2017p} was that the $q$-Laplace transform for 
$\l_N$ is written as a Fredholm determinant formula.
\begin{theorem}\label{ql-det}
For the two-sided $q$-Whittaker measure (\ref{2qWh-me}) with $0\leq \a_i < a_j \leq 1,1\leq i,j\leq N$ 
and with $\zeta\neq q^n,n\in\Z$, 
\begin{equation}
 \Big\langle \frac{1}{(\z q^{\l_N};q)_\i} \Big\rangle
 =
 \det(1-f K)_{L^2(\Z)},
 \label{det_formula}
\end{equation}
where $\langle \cdots \rangle$ means the average and   
\begin{align}
 f(n) &=  \frac{1}{1-q^n/\z}, \label{f}\\
 K(n,m) &= \sum_{l=0}^{N-1} \phi_l(m) \psi_l(n), \label{Kernel}\\
 \phi_l(n) 
 &=
 \int_D dv \frac{e^{-vt}}{v^{n+N}}\frac{1}{v-a_{l+1}}
 \prod_{j=1}^l \frac{v-\a_j}{v-a_j} \prod_{k=1}^N \frac{(q\a_k/v;q)_\i}{(qv/a_k;q)_\i} , 
 \label{phi}
\\
\psi_l(n)
&=
(a_{l+1}-\a_{l+1}) 
 \int_{C_r} dz \frac{e^{zt} z^{n+N}}{z-\a_{l+1}}
 \prod_{j=1}^l \frac{z-a_j}{z-\a_j} \prod_{k=1}^N \frac{(q z/a_k;q)_\i}{(q\a_k/z;q)_\i} . 
 \label{psi}
\end{align}
Here the contour $D$ is around $\{a_i,1\leq i\leq N\}$ 
and the contour $C_r$ is around $\{0,q^{i}\a_j,~i=0,1,2,\cdots,~1\leq j\leq N\}$.  
\end{theorem}

\section{Inverse Laplace transforms}
\label{sinvLap}
\subsection{Three versions} 
For a real function $\vp(x)$, defined for $x>0$, the Laplace transform is defined as 
\cite{Widder1941}
\begin{equation}
 \tilde{\vp}(u) = \int_0^{\i} e^{-ux} \vp(x) dx, ~ u\in\C. 
 \label{Lap}
\end{equation}
The region of $u$ in which the integral converges depends on $\vp(x)$. 
A formula to recover the original $\vp$ from its Laplace transform $\tilde{\vp}$ is well known,
\begin{equation}
 \vp(x) = \frac{1}{2\pi i} \int_{\d + i\R} du e^{ux} \tilde{\vp}(u),~x>0,
 \label{invLap}
\end{equation}
where $\d$ should be taken so that the singularities of $\tilde{\vp}$ are to the 
left of the integration contour. 

In this appendix,  we mainly consider the case where $\vp(x)$ is a (probability)
distribution function on $(0,\i)$. 
If a random variable having this distribution function $\vp(x)$ is denoted by $X$, 
its generating function $G(u)=\langle e^{-uX} \rangle$ is written as
\begin{equation}
 G(u) = \int_0^{\i} e^{-ux} d\vp(x) = u \int_0^{\i} e^{-ux} \vp(x) dx = u \tilde{\vp}(u).
 \label{GiL}
\end{equation} 
When $\vp(x)$ is a distribution function on $(0,\i)$,
$\tilde{\vp}(u)$ and hence also $G(u)$  
are analytic for ${\rm Re}\,u>0$. Hence in the inversion formula (\ref{invLap}), 
the condition on $\d$ is simply taken to be $\d>0$.

Here we discuss another inversion formula. 
\begin{proposition}
Let $\vp(x)$ be a distribution function on $(0,\i)$, decaying as $\vp(x) = O(x^a),a>0$
as $x$ approaches zero. Then we have 
\begin{equation}
 \vp(x) 
 = 
 \frac{1}{2\pi i} \int_{\d+i\R} d\xi \frac{x^{\xi}}{\Gamma(\xi+1)}
 \int_0^{\i} u^{\xi} \tilde{\vp}(u) du, ~x>0,
 \label{invLap2}
\end{equation}
where $\d>a$. 
\end{proposition}
\noindent
{\bf Remark.} 
Note that in (\ref{invLap2}) one needs only $\tilde{\vp}(u)$ for real $u>0$ 
to recover the original $\vp (x)$ whereas in (\ref{invLap})  the information of 
$\tilde{\vp}(u)$ for ${\rm Re }\,u > 0$ is necessary. 

\smallskip
\noindent
{\bf Proof.}
First we check rhs is finite. Since $\vp(x) =O(x^a)$ as $x\downarrow 0$, 
$\lim_{x\to\i}\vp(x)=1$ and ${\rm Re}\, \xi>a$,  the integral 
$\int_0^{\i} dx \frac{\vp(x)}{x^{\xi+1}}$ converges. (Note the integral 
does not converge when $\vp(x) = O(-1/\log x), x\downarrow 0$. )  
One then writes , for $\xi$ s.t. ${\rm Re}\, \xi>0$,
\begin{align}
 &\quad
 \Gamma(\xi+1) \int_0^{\i} dx \frac{\vp(x)}{x^{\xi+1}}
 =
 \int_0^{\i} dx \vp(x) \int_0^{\i} du u^{\xi} e^{-ux} \notag\\
 &=
 \int_0^{\i} du u^{\xi} \int_0^{\i} dx e^{-ux} \vp(x) 
 =
 \int_0^{\i} u^{\xi} \tilde{\vp}(u)du,
\end{align}
where for the second equality we used Fubini's theorem. Hence
\begin{align}
{\rm rhs~ of } \,(\ref{invLap2}) 
&=
\frac{1}{2\pi i} \int_{\d+i\R} d\xi x^{\xi} \int_0^{\i} dx_1 \frac{\vp(x_1)}{x_1^{\xi+1}}
=
\int_0^{\i} \frac{dx_1}{x_1} 
\vp(x_1) \frac{1}{2\pi i} \int_{\d+i\R} d\xi \left(\frac{x}{x_1}\right)^{\xi} \notag\\
&=
\int_{\R} dy_1 
\vp(e^{y_1}) \frac{1}{2\pi i} \int_{\d+i\R} d\xi e^{(y-y_1)\xi}
=
\vp(e^y) = \vp(x). 
\end{align}
\qed 

The Laplace transform $\tilde{\vp}(u)$ is often analytically continued to the region 
$\mathbb{C}\setminus \R_-$. One may find an analytic continuation directly from 
an expression for $\tilde{\vp}$. There is also a rather general lemma. 

\begin{lemma} (\cite{Ruijsenaars1997} App.~B)
Suppose $\vp(x)$ is analytic for ${\rm Re}\, x>0$ and at $x=0$,  and satisfies 
\begin{equation}
 |\vp(re^{i\th})| \leq C(\chi) , ~\forall (r,\th)\in [0,\i)\times [-\chi,\chi],
\end{equation}
where $\chi\in[0,\pi/2)$ and $C(\chi)$ is a positive non-decreasing function on $[0,\pi/2)$.  
Then $\tilde{\vp}(u)$ 
can be analytically continued to the region $\C\setminus \R_-$. 
\end{lemma}
\noindent
In such a case, we have the following third inversion formula. 
\begin{proposition} (\cite{BobylevCercignani2002})
Suppose $\tilde{\vp}(u)$ satisfies the followings. 

\begin{itemize}
\item[(i)] $\tilde{\vp}(u)$ is analytic on $\C\setminus\R_-$.

\item[(ii)] $\overline{\tilde{\vp}(u)}=\tilde{\vp}(\bar{u})$. 

\item[(iii)] The limiting value $\tilde{\vp}_{\pm}(t):=\tilde{\vp}(-t\pm i 0),t>0$ exist and 
     $\tilde{\vp}(t)=\overline{\tilde{\vp}(t)}$ holds.

\item[(iv)] $\tilde{\vp}(u) = o(1)$ for $|u|\to\i$ and $\tilde{\vp}(u) = o(1/|u|)$ for $|u|\to 0$,
uniformly in any sector $|{\rm arg} u|<\pi-\eta, \pi>\eta>0$. 

\item[(v)] There exists $\e>0$ s.t.  for every $\pi-\e < \phi \leq \pi$,
\begin{equation}
 \frac{\tilde{\vp}(r e^{\pm i\phi})}{1+r} \in L^1(\R_+), ~~ |\tilde{\vp}(r e^{\pm i\phi})| \leq a(r),
\end{equation}  
where $a(r)$ does not depend on $\phi$ and $a(r)e^{-\d r}\in L^1(\R_+)$ for any $\d>0$.  
\end{itemize}
Then 
\begin{equation}
 \vp(x) = \frac{1}{\pi} \int_0^{\i} e^{-xt }\, {\rm Im}\, \tilde{\vp}_-(t) dt. 
 \label{invLap3}
\end{equation}
\end{proposition}

\noindent
Basically the formula is obtained by changing the contour in (\ref{invLap}) to the one 
around $\R_-$ and then take the limit to $\R_-$ from both above and below. 

\vspace{5mm}
Suppose $F(y)$ is a distribution function on $\R$ associated with a random variable $Y$. 
Then $\vp(x)=F(\log x),~x >0$ is a distribution function on $(0,\i)$ and the above formulas 
can be applied.  
First, combining (\ref{invLap}) and (\ref{GiL}) with $x=e^y$,  we have 
\begin{equation}
 F(y) = \frac{1}{2\pi i} \int_{\d + i\R} \frac{du}{u} e^{ue^y} G(u),
 \label{invLapF1}
\end{equation}
where $G(u)$ is written in terms of $Y$ as $G(u) = \langle e^{-ue^Y}\rangle$.
Next (\ref{invLap2}) is rewritten  as follows. 
\begin{corollary}
For a random variable $Y$, set $G(u) = \langle e^{-ue^Y}\rangle$. 
The distribution function of $Y$ is recovered from $G(u)$ as
\begin{equation}
 F(y) 
 = 
 \frac{1}{2\pi i} \int_{\d+i\R} d\xi \frac{e^{y \xi}}{\Gamma(\xi+1)}
 \int_0^{\i} u^{\xi-1} G(u)du. 
 \label{GinvLap2F}
\end{equation}
\end{corollary} 
\noindent
The corresponding density function $f(y)=F'(y)$, if it exits, is given by 
\begin{equation}
 f(y) 
 = 
 \frac{1}{2\pi i} \int_{\d+i\R} d\xi \frac{e^{y \xi}}{\Gamma(\xi)}
 \int_0^{\i} u^{\xi-1} G(u) du. 
 \label{invLap2f}
\end{equation}
There is an analogous formula for general $n$th derivative, when they exist. 
Note 
(\ref{GinvLap2F})  is equivalent to 
\begin{equation}
 F^{\sharp}(\xi) 
 := 
 \int_\R dy e^{-y\xi} F(y) 
 =
 \frac{1}{\Gamma(\xi+1)} \int_0^{\i} u^{\xi-1} G(u) du,
  \label{invLap2FF}
\end{equation}
where $\xi\in i\R + \delta, \delta>0$. For $\delta\to 0$, this is the Fourier transform. 

The third Laplace inversion formula (\ref{invLap3}) reads
\begin{equation}
 F(y) 
 = 
 1-\int_\R dw e^{-e^{y-w}} \frac{1}{\pi} \lim_{\e\downarrow 0}{\rm Im}\, G(-e^{-w}+i\e) . 
 \label{inv3}
\end{equation}
One notices that this is written in a form of convolution including the Gumbel distribution. 
In the context of the KPZ equation, the distribution function in this form 
appeared in \cite{SS2010b}
and an explanation of the appearance of the Gumbel distribution was given in \cite{CLDR2010}. 
An advantage of this inversion formula as compared to (\ref{GinvLap2F}) 
is that whereas the latter still contains the complex integral over $\xi$, 
in the former all quantities are real. This is a useful  property, for 
example when evaluating the numerical value of the distribution function. 

\subsection{Sum of two independent random variables}
Let us consider the case in which a random variable $Y$ can be written in a form, 
\begin{equation}
Y = Y_0+\chi,
\end{equation}
where $Y_0$ and $\chi$ are independent. We discuss a few formulas which 
give the distribution $F_0(y):=\P[Y_0\leq y]$ in terms of the information on $Y$ and $\chi$. 
The discussions in this subsection are rather formal. We simply assume that all the 
quantities have appropriate analytic properties. 
A most standard approach would be to consider the Fourier transforms. 
By the independence, the Fourier transforms, $F^{\sharp}(\xi), F_0^{\sharp}(\xi),g(\xi)$, 
are related as 
\begin{equation}
 F^{\sharp}(\xi) = g(\xi) F_0^{\sharp}(\xi). 
\label{FgF0}
\end{equation}
Hence the distribution function $F_0$ can be written as 
\begin{equation}
 F_0(y) = \int d\xi e^{y\xi} F_0^{\sharp}(\xi) 
          = \int d\xi \frac{e^{y\xi} F^{\sharp}(\xi)}{g(\xi)}. 
\end{equation}
The formula (\ref{lp13}) is a result of the combination of this and 
(\ref{GinvLap2F}). 
For the distributions themselves, the independence implies
\begin{equation}
 F(y)  = g\left(\frac{d}{dy}\right) F_0(y). 
\end{equation}
Note that (\ref{FgF0}) is the Fourier transform of this relation. 
By inverting this, we find 
\begin{equation}
 F_0(y) = \frac{1}{g\left(\frac{d}{dy}\right)} F(y). 
\end{equation}
This may look a formal expression, but when $1/g(\xi)$ can be 
Taylor expanded and the resulting series
\begin{equation}
 \sum_{n=0}^{\i}\frac{1}{n!} \frac{d^n}{dy^n}\left(\frac{1}{g(y)}\right)\Bigr|_{y=0} \cdot F^{(n)}(y)
\end{equation}
converges, this makes sense. The formula (2.22) in \cite{IS2013}
is a result of the combination of this and (\ref{inv3}).

We can also discuss similar relations for the generating functions. 
Let us set $Z=e^{Y}, Z_0=e^{Y_0}$ and  
\begin{equation}
 G(u) = \langle e^{-uZ} \rangle ,~
 G_0(u)=\langle e^{-uZ_0} \rangle ,~
 g(u) = \langle e^{u e^{\chi}} \rangle.
\end{equation} 
By the independence, we have 
\begin{equation}
 G(e^v)  = g\left(\frac{d}{dv}\right) G_0(e^v) . 
 \label{GgG0}
\end{equation} 
By inverting this, we find
\begin{equation}
 G_0(e^v) = \frac{1}{g\left(\frac{d}{dv}\right)} G(e^v). 
 \label{G0gG}
\end{equation}

\section{Proof of Lemma~\ref{ll7}}
\label{proofll7}
First we consider~\eqref{ll71} by the saddle point analysis, especially 
$\psi_l(\tilde{x})$~\eqref{l48} since we can deal with $\phi_l(\tilde{x})$~\eqref{l47} 
in a parallel way.
Changing $w=-i\sqrt{N}z$ in~\eqref{l48} , one has
\begin{align}
\psi_l(\tilde{x})=\frac{\sqrt{N}}{2\pi i}\int_{-i\infty}^{i\infty}
dz\, e^{f_N(z;\t,\tilde{x},l)}
\frac
{\Gamma (1+\sqrt{N}z)\Gamma \left(1+\a-\sqrt{N}z\right)}
{\Gamma \left(1-\a+\sqrt{N}z\right)}
,
\label{psifN}
\end{align}
where
\begin{align}
f_N(z;\t,\tilde{x},l)=-\sqrt{N}z\tilde{x}+N\frac{z^2t}{2}+(l-N)\log(\sqrt{N}z)
-N\log\Gamma (\sqrt{N} z).
\label{fN}
\end{align}
Substituting~\eqref{l31} with~\eqref{l200} and~\eqref{ll76} into~\eqref{fN}, 
we find that the first three terms of~\eqref{fN} becomes
\begin{align}
&~-\sqrt{N}z\tilde{x}+N\frac{z^2\t}{2}+(l-N)\log(\sqrt{N}z)\notag\\
&=N^{3/2}\log N \cdot\frac{z}{2}
+N^{3/2}\left(-z+\frac{t^{1/2}z^2}{2}-\frac{z}{2}\log t\right)
+N\left(\gamma_t^2yz^2-2\gamma_t^2yt^{-1/2}z-\frac{t^{1/2}z}{2}\right)\notag\\
&~~-N^{1/2}\log N\cdot\frac{\lambda}{2(2\gamma_t)^{1/2}} 
+N^{1/2}\left(\frac{\gamma_t^3}{12}z-\gamma_t\xi_i z
+\gamma_ty^2z-\gamma_t^2yz
-\frac{\lambda}{(2\gamma_t)^{1/2}}\log z\right).
\label{fN123}
\end{align}
For the last term we use the Stirling formula,
\begin{align}
\log\Gamma(n)=n\log n-n-\frac{\log2\pi n}{2}+\frac{1}{12n}+O(n^{-3})
\label{ll718}
\end{align}
and have
\begin{align}
-N\log\Gamma (\sqrt{N}z)
&=-N^{3/2}\log N\cdot \frac{z}{2}+
N^{3/2}\left(z-z\log z\right)+\frac{N}{4}\log N+
N \frac{\log2\pi z}{2}\notag\\
&~~~~-N^{1/2}\frac{1}{12z}+O(N^{-1/2}).
\label{fN4}
\end{align}
Using~\eqref{fN123} and~\eqref{fN4},  we write $f_N(z;\t,\tilde{x},l)$~\eqref{fN} 
as
\begin{align}
&f_N(z;\t,\tilde{x},l)=N^{3/2}f_{1}(z)+Nf_{2}(z)+N^{1/2}f_{3}(z)+C_1+O(N^{-1/2}),
\label{fNex}
\\
&f_{1}(z)=\frac{t^{1/2}z^2}{2}-z\log z-\frac{z}{2}\log t,
\label{fz}
\\
&f_{2}(z)=-\frac{t^{1/2}z}{2}+\frac{\log z}{2}+\gamma_t^2yz^2-\frac{2\gamma_t^2y}{y^{1/2}}z,
\label{gz}
\\
&f_{3}(z)=-\frac{1}{12z}+\left(\frac{\gamma_t^3}{12}-\gamma_t^2y-\gamma_t(\xi_i-y^2)
\right)z
-\frac{\lambda}{(2\gamma_t)^{1/2}}\log z.
\label{hz}
\end{align}
Here $C_1(N)$ is a constant , which does not depend on $z$,
$
C_1=-N^{1/2}\log N\cdot\frac{\lambda}{2(2\gamma_t)^{1/2}}+\frac{N}{2}\log 2\pi
\sqrt{N}.
$
We note that $f_{1}(z)$ above has a double saddle point
$z_c=t^{-1/2}$ such that $f_{1}'(z_c)=f_{N,1}''(z_c)=0$. We expand 
$f_{1}(z),~f_{2}(z),~f_{3}(z)$ around $z_c$. Noting
$f'''_{1}(z_c)=2\gamma_t^3$, $f'_{2}        (z_c)=0,~f_{2}''(z_c)=2\gamma_t^2y-\gamma_t^3$, $f_{3}'(z_c)=\gamma_t^3/4-\gamma_t^2y-\gamma_t(\lambda+\xi-y^2)$,
we get
\begin{align}
N^{3/2}f_{1}(z)
&=N^{3/2}f_{1}(z_c)+N^{3/2}\frac{\gamma_t^3}{3}(z-z_c)^3+O(N^{3/2}(z-z_c)^4),
\label{fex}
\\
Nf_{2}(z)
&=Nf_{2}(z_c)-N\left(\frac{\gamma_t^3}{2}-\gamma_t^2y\right)(z-z_c)^2+O(N(z-z_c)^3),
\label{gex}
\\
N^{1/2}f_{3}(z)
&=N^{1/2}f_{3}(z_c)+N^{1/2}\left(\frac{\gamma_t^3}{4}-\gamma_t^2y-\gamma_t(\lambda+\xi-y^2)
\right)(z-z_c)+O(N^{1/2}(z-z_c)^2).
\label{hex}
\end{align}
Here we scale $z$ around the critical point $z_c=1/\sqrt{t}$
\begin{align}
z=z_c+\frac{1}{\sqrt{N}}
\left(
\frac{1}{2}-\frac{1}{\gamma_t}(y+i\sigma)
\right).
\label{ll76d}
\end{align}
Using~\eqref{fNex}--\eqref{hex} with the fact
\begin{align}
f_N(\tilde{z};\t,\tilde{x},l)=C_1+C_2+N^{3/2}f_{1}(z_c)+N f_{2}(z_c)+N^{1/2}f_{3}(z_c)+O(N^{-1/2}),
\label{l111}
\end{align}
where $\tilde{z}$ is defined as~\eqref{ll76d} with $\sigma=0$, $C_1$
is defined below~\eqref{hz} and
\begin{align}
C_2=-\frac{1}{3}\left(y-\frac{\gamma_t}{2}\right)^3+\left(\xi-\lambda\right)
\left(y-\frac{\gamma_t}{2}\right),
\label{C5}
\end{align}
we have
\begin{align}
f_N(z;\t,\tilde{x},l)=\frac{i}{3}\sigma^3+i(\xi+\l)\sigma
+f_N(\tilde{z};\t,\tilde{x},l)+O\left(N^{-1/2}\right).
\label{ll77}
\end{align}

At last we evaluate the remaining factor with the Gamma functions 
in~\eqref{psifN}. Substituting~\eqref{ll76}, we have
\begin{align}
\frac
{\Gamma (1+\sqrt{N}z)\Gamma \left(1+\a-\sqrt{N}z\right)}
{\Gamma \left(1-\a+\sqrt{N}z\right)}
=
\Gamma \left(\frac32-\frac{y+i\sigma}{\gamma_t}+\sqrt{\frac{N}{t}}\right)
\frac
{\Gamma\left(1+(i\sigma+y+\o)/\g_t\right)}
{\Gamma\left(1-(i\sigma+y+\o)/\g_t\right)}.
\label{ll710}
\end{align}

Thus from~\eqref{psifN} and \eqref{ll77}, we find that
the limiting form of $\psi_l(\tilde{x})$ becomes 
\begin{align}
\lim_{N\rightarrow\infty}\frac{e^{f_N(\tilde{z};\t,\tilde{x},l)}}{\Gamma\left(1+\sqrt{N}\tilde{z}\right)} \psi_l(\tilde{x})
&=
\frac{1}{2\pi}\int_{-\infty}^{\infty}d\sigma\,e^{\frac{i}{3}\sigma^3+i(\xi_i+\lambda)\sigma}
\frac
{\Gamma (1+(i \sigma+\o+y)/\gamma_t)}
{\Gamma (1-(i \sigma+\o+y)/\gamma_t)}
\notag
\\
&=\Ai_{\Gamma}(\xi+\lambda,\o+y),
\label{ll78}
\end{align}
which is the second relation in~\eqref{ll71}.

We can show the result of $\phi_l(\tilde{x})$ in~\eqref{ll71}
in a parallel way. By changing $v=\sqrt{N}z$, 
$\phi_l(\tilde{x};t)$~\eqref{l47} is rewritten as
\begin{align}
\phi_l(\tilde{x})=\frac{1}{2\pi i}\oint dz\, \frac{e^{-f_N(z;\t,\tilde{x},l)}}{z}
\frac
{\Gamma \left(1-\a+\sqrt{N}\right)}
{\Gamma (1+\sqrt{N}z)\Gamma \left(1+\a-\sqrt{N}\right)},
\end{align}
where $f_N(z;\t,x,l)$ is given in~\eqref{fN}.  Applying the same techniques as 
the case of $\psi_l(x)$, we get the first relation in~\eqref{ll71}. 

Next we derive~\eqref{ll72}. As in the case of~\eqref{ll71}, we mainly
consider $B_{2}^{(1)}(\tilde{x})$~\eqref{l25}. From the definition of $f_N(z;\t,\tilde{x},k)$~\eqref{fN},  
it can be written as
\begin{align}
\tilde{B}_{2}^{(1)}(\tilde{x})
=
e^{f_N(\a/\sqrt{N},\t,\tilde{x},N-1)}
\Gamma(1+\a).
\label{ll711}
\end{align}
Note that from~\eqref{l31}, $\a/\sqrt{N}$ is scaled as 
$
\frac{\a}{\sqrt{N}}
=z_c
+\left(\frac{1}{2}+\frac{\o}{\g_t}\right)\frac{1}{\sqrt{N}}
$
with $z_c=1/\sqrt{t}$. Comparing this with~\eqref{ll76d},
$f_N(\a/\sqrt{N},\t,\tilde{x},N-1)$ in~\eqref{ll711} can be estimated by~\eqref{ll77}
with $\l=0$ and $\sigma=i(v+y)$ leading to
\begin{align}
&f_N(\a/\sqrt{N},\t,\tilde{x},N-1)
\notag
\\
&=
\frac{1}{3}(\o+y)^3-\xi (\o+y)+f_N\left(z_c+\frac{1}{2\sqrt{N}};\t,\tilde{x},N-1\right)
+O\left(N^{-\frac12}\right).
\label{ll713}
\end{align}
The second part of~\eqref{ll72} follows immediately from~\eqref{ll711}
and~\eqref{ll713}. We can also obtain the first part in a similar way.

Third we derive~\eqref{ll73}. We mainly consider the second relation.
As with the case~\eqref{ll71}, by the change of the variable
$w=-i\sqrt{N}z$,
$B^{(2)}_2(\tilde{x})$~\eqref{l27} can be expressed as
\begin{align}
\tilde{B}^{(2)}_2(\tilde{x})=\frac{\sqrt{N}}{2\pi i}\int_{-i\R+c}
dz\, 
\frac{e^{f_N(z;\t,\tilde{x},N-1)}}
{\a-\sqrt{N}z}
\frac
{\Gamma (1+\sqrt{N}z)\Gamma \left(1+\a-\sqrt{N}z\right)}
{\Gamma \left(1-\a+\sqrt{N}z\right)}
\label{ll79}
\end{align}
with $\a/\sqrt{N}<c<(\a+1)/\sqrt{N}$. 
Thus we get
\begin{align}
\lim_{N\rightarrow\i}\frac{e^{-f_N(z_c;\t,\tilde{x},N-1)}}{\Gamma\left(1+\sqrt{N/t}\right)} 
\tilde{B}_2^{(2)}(\tilde{x})
=
\frac{1}{2\pi}\int_{\R+ic}d\sigma\,
\frac
{e^{\frac{i}{3}\sigma^3+i(\xi_i-\lambda)\sigma}}
{y+\o+i\sigma}
\frac
{\Gamma (1+(i\sigma+\o+y)/\gamma_t)}
{\Gamma (1-(i\sigma+\o+y)/\gamma_t)}
\label{ll80}
\end{align}
with $y+\o<c<y+\o+\g_t$. Using 
the relation
\begin{align}
\frac{1}{y+\o+i\sigma}=-\int_{0}^{\infty}d\l
e^{\l(y+\o+i\sigma)},
\end{align}
which is confirmed by $\text{Re}(y+\o+i\sigma)<0$,
we arrive at the second relation of~\eqref{ll73}.
The first relation can also be shown in the same way. 

At last we consider~\eqref{ll74}. Taking the scalings~\eqref{l31} with~\eqref{l200} and~\eqref{ll75}
into account, we see that each term in lhs of~\eqref{ll74} becomes
\begin{align}
&(N-1)
\frac{\Gamma'(1+\a)}{\Gamma(1+\a)}
=
\frac12 (N-1)\log\frac{N}{t}+\sqrt{Nt}\left(1+\frac{\o}{\gamma_t}\right)
-t\left(\frac{11}{24}+\frac{\o}{\gamma_t}+\frac{\o^2}{2\gamma_t^2}\right)
+O(N^{-\frac12}),
\label{ll714}
\\
&-\frac{N-1}{\a}
=
-\sqrt{Nt}+t\left(\frac{1}{2}+\frac{\o}{\gamma_t}\right)+O(N^{-\frac12}),
\label{ll715}
\\
&-\a\t
=
-N+\sqrt{N}\left(-\frac{\sqrt{t}}{2}-\frac{\sqrt{t}\o}{\gamma_t}
-\frac{2\gamma_t^2y}{\sqrt{t}}
\right)
-2\gamma_t^2y\left(\frac{1}{2}+\frac{\o}{\gamma_t}\right)
+O(N^{-\frac12}),
\label{ll716}
\\
&-\log\tilde{u}
=
-\frac{N-1}{2}\log\left(\frac{N}{t}\right)
+
N
+
\sqrt{N}\left(\frac{\sqrt{t}}{2}+\frac{2\gamma_t y}{\sqrt{t}}\right)
\notag
\\
&
\hspace{6cm}
-\frac{\gamma_t^3}{12}+\gamma_t^2 y
+
\gamma_t (s-y^2)
-\log u
+
O(N^{-\frac12}),
\label{ll717}
\end{align}
where in~\eqref{ll714}, we used~\eqref{ll718}.
\eqref{ll74} follows immediately from~\eqref{ll714}--\eqref{ll717}.
\qed

\providecommand{\bysame}{\leavevmode\hbox to3em{\hrulefill}\thinspace}
\providecommand{\MR}{\relax\ifhmode\unskip\space\fi MR }
\providecommand{\MRhref}[2]{%
  \href{http://www.ams.org/mathscinet-getitem?mr=#1}{#2}
}
\providecommand{\href}[2]{#2}

\end{document}